\documentclass{article}
\pdfoutput=1
\usepackage{arxiv}
\usepackage[utf8]{inputenc} % allow utf-8 input
\usepackage[T1]{fontenc}    % use 8-bit T1 fonts
\usepackage{hyperref}       % hyperlinks
\usepackage{url}            % simple URL typesetting
\usepackage{booktabs}       % professional-quality tables
\usepackage{amsfonts}       % blackboard math symbols
\usepackage{nicefrac}       % compact symbols for 1/2, etc.
\usepackage{microtype}      % microtypography
\usepackage{lipsum}
\usepackage{graphicx}
\usepackage{booktabs}
\usepackage{booktabs,threeparttable}
\usepackage{array}
\usepackage{paralist}
\usepackage{verbatim}
\usepackage{subfig}
\usepackage{amsmath}
\usepackage{tikz}
\usepackage{pgfplots}
\usepackage{graphicx}
\usepackage{gensymb}
\pgfplotsset{width=10cm ,compat=1.9}
\usepackage{listings}
\usepackage{xcolor}
\usepackage{fancybox}
\usepackage{float}
\usepackage{url}
\usepackage{extpfeil}
\usepackage[mathscr]{euscript}
\usepackage{hyperref}

\newcommand\blfootnote[1]{%
	\begingroup
	\renewcommand\thefootnote{}\footnote{#1}%
	\addtocounter{footnote}{-1}%
	\endgroup
}

\title{Shared Control Between Pilots and Autopilots: Illustration of a Cyber-Physical Human System}

\author{
  Emre Eraslan\\
  Department of Mechanical Engineering\\
  Bilkent University\\
  \texttt{emre.eraslan@bilkent.edu.tr} \\
  \And
  Yildiray Yildiz \\
  Department of Mechanical Engineering\\
  Bilkent University\\
  \texttt{yyildiz@bilkent.edu.tr} \\
  \AND
  Anuradha M.~Annaswamy\\
  Department of Mechanical Engineering\\
  Massachusetts Institute of Technology\\
  \texttt{aanna@mit.edu} \\
}

\begin{document}
\maketitle

\begin{abstract}
Although increased automation has made it easier to control aircraft, ensuring a safe interaction between the pilots and the autopilots is still a challenging problem, especially in the presence of severe anomalies. Current approach consists of autopilot solutions that disengage themselves when they become ineffective. This may cause reengagement of the pilot at the worst possible time, which can result in undesired consequences. In this paper, a series of research studies that propose pilot-autopilot interaction schemes based on the Capacity for Maneuver (CfM) concept, are covered. CfM refers to the remaining capacity of the actuators that can be used for bringing the aircraft to safety. It is demonstrated that CfM-based pilot-autopilot interaction schemes, or \textit{Shared Control Architectures} (SCA), can be promising alternatives to the existing schemes. Two SCAs are tested in the experiments. In SCA1, the pilot takes over the control from the autopilot using the monitored CfM information. In SCA2, the pilot takes on the role of a supervisor helping the autopilot whenever a need arises, based on the CfM information. Whenever the aircraft experiences a severe anomaly, the pilot assesses the situation based on his/her CfM monitoring and intervenes by providing two control system parameter estimates to the autopilot. This increases the effectiveness of the autopilot. Using human-in-the-loop simulations, it is shown that CfM based interactions provides smaller tracking errors and larger overall CfM. The subjects including a commercial airline pilot and several university students are trained using a systematic procedure. Creation of new cyber physical \& human systems is inevitable along with deeper engagement with the social science community so as to get better insight into human decision making. The results reported here should be viewed as a first of several steps in this research direction.  \blfootnote{“© 2019 IEEE.  Personal use of this material is permitted.  Permission from IEEE must be obtained for all other uses, in any current or future media, including reprinting/republishing this material for advertising or promotional purposes, creating new collective works, for resale or redistribution to servers or lists, or reuse of any copyrighted component of this work in other works.”}

\end{abstract}

% keywords can be removed

\section{Introduction}

The 21st century is witnessing large transformations in several sectors including energy,
transportation, robotics, and health care related to autonomy. Decision-making using real-time information over a large range of operations as well as the ability to adapt online in the presence of various uncertainties and anomalies is the hallmark of an autonomous system. In order to design such a system, a variety of challenges needs to be addressed. Uncertainties may occur in several forms, both structured and unstructured. Anomalies may often be severe that require rapid detection and swift action to minimize damage and restore normalcy. This paper addresses the difficult task of making autonomous decisions in the presence of severe anomalies. While the specific application we focus on is flight control, the overall solutions we propose are applicable for general complex dynamic systems.

The domain of decision making is common to both human experts and feedback control systems. Human experts routinely make several decisions when faced with anomalous situations. In the specific context of flight control systems, pilots often take several decisions based on the sensory information from the cockpit, situational awareness, their expert knowledge of the aircraft, and ensure a safe performance. All autopilots in a fly-by-wire aircraft are programmed to provide the appropriate corrective input to ensure that the requisite variables follow the specified guidance commands accurately. Advanced autopilots ensure that such a command following occurs even in the presence of uncertainties and anomalies. However, the process of assembling various information that may help detect the anomaly may vary between the pilot and the autopilot. Once the anomaly is detected, the process of mitigating the impact of the anomaly may also differ between the pilot and autopilot. Nature of perception, speed of response, intrinsic latencies may all vary significantly between the two decision makers. It may be argued that perception and detection of the anomaly may be carried out efficiently by the human pilot whereas fast action following a command specification may be best accomplished by an autopilot. Our thesis in this paper is that a shared control architecture where the decision making of the human pilot is judiciously combined with that of an advanced autopilot is highly attractive in flight control problems where severe anomalies are present.

\begin{figure}[htb]
	\centering
	\includegraphics[scale=0.70]{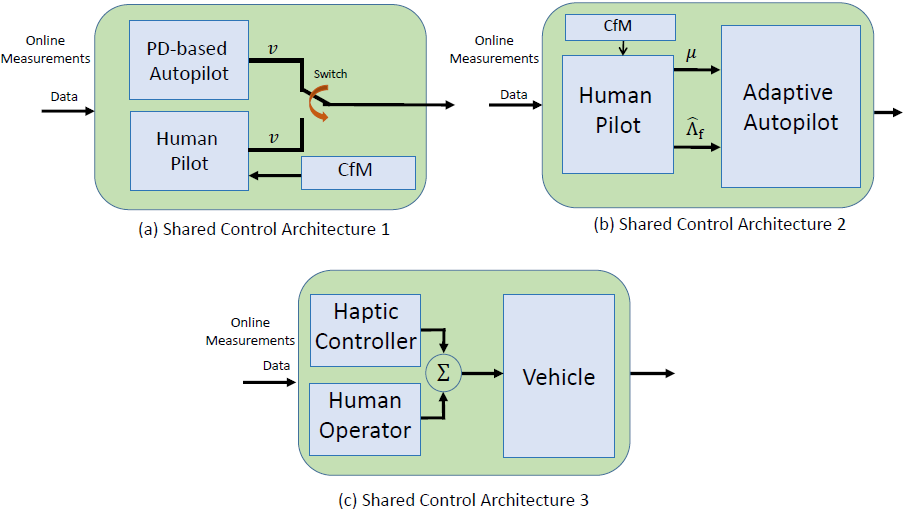}
	
	\caption{An overview of Shared Control Architectures (SCAs). (a) represents SCA1, in which the autopilot takes care of the control task until an anomaly, which is then transferred to the pilot based on Capacity for Maneuver (CfM). (b) represents the SCA2, in which the pilot undertakes a supervisory role. CfM values approaching to dangerously small numbers may trigger the pilot to assess the situation and convey necessary auxiliary inputs $\mu$ and $\hat \Lambda_{f_p}$ to the autopilot. (c) represents SCA3, which is based on the combined control effort continuously transmitted by both haptic controller and human operator who has the choice of abiding by or dismissing the control support of the automation.} 
	\label{fig:scas_overall_view}
\end{figure}

Existing frameworks that combine humans and automation generally rely on human experts supplementing flight control automation, i.e., the system is semi-autonomous, with the automation doing all of the work to handle uncertainties and disturbances, and human taking over control once the environment imposes demands that exceeds the automation’s capabilities \cite{hess2009modeling, hess2014model, hess2016modeling}. This approach causes bumpy and late transfer of control from the machine to the human, causing the shared-control architecture to fail, as it is unable to keep pace with the cascading demand which may cause actual accidents \cite{woods2006joint, woods2011basic, icing96, woods2018theory}. This  suggests that alternate architectures of coordination and interaction between the human expert and automation may be needed. Various such architectures have been proposed in the literature over the years, which can be broadly classified into three forms: a trading action where humans take over control from automation under emergency conditions \cite{abbink2018topology}, a supervisory action where the pilot assumes a high-level role and provides the inputs and setpoints for the automation to track \cite{sheridan2013monitoring}, and  a combined action where both the automation and the human expert participate at the same time scale \cite{mulder2012sharing}. We label all  three forms under a collective name of Shared Control Architecture (SCA), with SCA1 denoting the trading action, SCA2 denoting the supervisory action, and SCA3 denoting the combined action (see Figure \ref{fig:scas_overall_view} for a schematic of all three forms of SCAs). In this paper, we focus on SCA1 and SCA2, and their validation using human-in-the-loop simulations. 

In order to have an efficient coordination between humans and automation, we utilize two principles from cognitive engineering that studies how humans add resilience to complex systems \cite{woods2006joint, woods2011basic, woods2018theory}. The first principle is Capacity for Maneuver (CfM), which denotes a system’s reserve capacity that will remain after the occurrence of an anomaly \cite{woods2011basic}. It is hypothesized that resiliency is rooted in and achieved via monitoring and regulation of the system’s CfM \cite{woods2018theory}. The notion of CfM exists in an engineering context as well, which is the reserve capacity of an actuator. Viewing the actuator input power as the system’s capacity, and noting that a fundamental capacity limit exists in all actuators in the form of magnitude saturation, one can define CfM for an autopilot-controlled system as the distance between the control input and saturation limits. The need to avoid actuator saturation, and therefore increasing CfM, becomes even more urgent in the face of anomalies which may push the actuators to their limits. This implies that there is a common link between a pilot-based decision making with that of an autopilot-based one in the form of CfM. We utilize this commonality in both the SCA1 and SCA2 presented in this paper.

The foundations for SCA1 and SCA2 come from the results of \cite{farjadian2016resilient, thomsen2019shared, farjadian2018resilient}, with \cite{farjadian2016resilient} and \cite{thomsen2019shared} corresponding to SCA1 and \cite{farjadian2018resilient} corresponding to SCA2. In \cite{farjadian2016resilient}, the assumption is that the autopilot is designed to accommodate satisfactory operation under nominal conditions, with the requisite tracking performance. An anomaly is assumed to occur in the form of loss of effectiveness of the actuator (which may be due to a damage to the control surfaces caused by a sudden change in environmental conditions or a compromised engine due to bird strikes). The trading action proposed in \cite{farjadian2016resilient} is for the pilot to step in and take over from the autopilot, based on the pilot’s perception that the automation is unable to cope with the anomaly. This perception is based on the CfM of the actuator, and when it exceeds a certain threshold, the pilot is proposed to take over control.  A well known model of the pilot in \cite{hess2009modeling, hess2014model, hess2016modeling} is utilized to propose the specific sequence of control actions that the pilot takes once they take over. It is shown through simulation studies that when the pilot carries out this sequence of perception and control tasks, the effect of anomaly is contained, and the tracking performance is maintained at a satisfactory level. A slight variation of the above trading role is reported in \cite{thomsen2019shared}, where instead of the ``autopilot is active$\rightarrow$anomaly occurs$\rightarrow$pilot takes over'' sequence, the pilot transfers control from one autopilot to a more advanced autopilot following the perception of an anomaly. In this paper, we limit our attention to the SCA1 architecture in \cite{farjadian2016resilient} and validate its performance using human-in-the-loop simulations.

The SCA2 architecture utilized in \cite{farjadian2018resilient} differs from SCA1, as mentioned above, as the trading action from the autopilot to the pilot is replaced by a supervisory action by the pilot. In addition to using CfM, this architecture utilizes a second principle from cognitive engineering denoted as Graceful Command Degradation (GCD). GCD is proposed as an inherent metric adopted by humans \cite{woods2006joint} that will allow the underlying system to function so as to retain a target CfM. As the name connotes, GCD corresponds to the extent to which the system is allowed to relax its performance goals. When subjected to anomalies, it is reasonable to impose a  command degradation; the greater this degradation, the larger the reserves that the system will possess when recovered from the anomaly. A GCD can then be viewed as a control variable that is tuned so as to permit a system to reach its targeted CfM. The role of tuning this variable so that the desired CfM is retained by the system is relegated to the pilot in \cite{farjadian2018resilient}. In particular a  parameter $\mu$ is transferred to the autopilot from the autopilot, which is shown to result in an ideal trade-off between the CfM and GCD by the overall closed-loop system with SCA2. In this paper, we validate SCA2 with a high-fidelity model of an aircraft and with human-in-the-loop simulations. 

The subjects used in the human-in-the-loop simulations include an airline pilot, who is also a flight instructor, with 2600 hours of flight experience, as well as human subjects who were trained in a systematic manner by the experiment designer. The details of the training is explained in the Section “Validation with Human in the Loop Simulations”.  It is shown that the SCA1 in \cite{farjadian2016resilient} and SCA2 in \cite{farjadian2018resilient} indeed lead to better performance as conjectured therein when the human expert carries out their assigned roles in the respective architectures. The resulting solution therefore is an embodiment of an efficient cyber-physical human system, a topic that is of significant interest of late.

In addition to the references mentioned above, SCAs have been explored in  haptic shared control (HSC), which has applications both in the aerospace \cite{smisek2016neuromuscular} and in automotive \cite{van2015driver, mulder2012sharing} domains. In HSC, both the automation and the human operator exert forces on a common control interface, such as a steering wheel or a joystick, to achieve their individual goals. In this regard, HSC can be considered to have a SCA3 architecture. In \cite{smisek2016neuromuscular}, the goal of the haptic feedback is to help a UAV operator avoid an obstacle. In \cite{van2015driver}, one of the goals is lane-keeping and in \cite{mulder2012sharing}, the goal is assisting the driver while driving around a curve. All of these approaches provide situational awareness to the human  operator and the human has the ability to override the haptic feedback by exerting more force to the control interface. An interesting HSC study is presented in \cite{smisek2014adapting} where the haptic guidance authority is modified adaptively according to the intensity of the user grip, in which, challenging scenarios are created by introducing force disturbances or incorrect guidance. Another SCA approach is to enable the automation to affect the control input directly instead of using a common interface \cite{rossetter2006lyapunov}. In this approach, the human operator can override the automation by disengaging it from the control system. Since SCA1 and SCA2 architectures help with anomaly mitigation and SCA3 is not shown to be relevant in this respect, in this paper, we mainly focus on SCA1 and SCA2 and their validations with human-in-the-loop simulation studies. 

%Are these references SCA-3? If so, we can justify as to why we only look at SCA-1 and SCA-2 as they help with anomaly-mitigation whereas SCA-3 has not been shown to be relevant for this goal?

In the specific area of flight control, SCAs have also been proposed in order to lead to greater situational awareness. A bumpless transfer between the autopilot and pilot is proposed to occur in \cite{ackerman2015flight} using a more informative pilot display. It is argued in \cite{ackerman2015flight} that a common problem in pilot-automation interaction is the unawareness of the pilot of the automation state and the aircraft during the flight, which makes the automation system opaque to the pilot, prompting the use of such a pilot display. The employment of the display is discussed in \cite{ackerman2015flight}, in the presence of automation developed in \cite{tekles2014flight, chongvisal2014loss} that provides flight envelope protection together with a loss of control logic. A review of recent advances on human-machine interaction can be found in \cite{Kaber:2013}, where the focus is on adaptive automation, in which the automation monitors the human pilot to adjust itself accordingly, so as to lead to improved situational awareness. No severe anomalies are however considered in \cite{ackerman2015flight, tekles2014flight, chongvisal2014loss, Kaber:2013} which is the focus of this paper. The goal of the shared control architectures discussed in this article consisting of the human and automation is to lead to a bumpless efficient performance in the presence of severe anomalies.

In the following sections, we first introduce the two main components of SCA, which are  the autopilot and the human pilot. In the “Autopilot” section, we provide the details of the employed controllers. In the “Human Pilot” section, we first review main mathematical models of pilots proposed in the literature, in the absence of a severe anomaly. Then, we present the human model used in the development of the proposed SCAs. We later present the working principles of these SCAs in detail, in section “Shared Control”. Finally, in the “Validation with Human in the Loop Simulations” section, we show how the underlying ideas are tested with human subjects.

\section{Autopilot}

In this section, we describe the technical details related to the autopilot discussed in \cite{farjadian2016resilient} and \cite{farjadian2018resilient}.  
\subsection{Dynamic Model of the Aircraft}
Since the autopilot in \cite{farjadian2018resilient} is assumed to be determined using feedback control, the starting point is the description of the aircraft model, which has the form
\begin{equation}\label{e:plant}
\dot{x}(t)=Ax(t)+B\Lambda_f u(t)+d+\Phi^T f(x),
\end{equation}
where $x\in\mathbb{R}^n$ and $u\in\mathbb{R}^m$ are deviations around a trim condition in aircraft states and control input, respectively, $d$ represents uncertainties associated with the trim condition, and the last term $\Phi^T f(x)$ represents higher order effects due to nonlinearities. $A$ is a $(n\times n)$ system matrix and $B$ is a $(n\times m)$ input matrix, both of which are assumed to be known, with $(A,B)$ controllable, and $\Lambda_f$ is a diagonal matrix that reflects a possible actuator anomaly with unknown positive entries $\lambda_{f_i}$. It is assumed that the anomalies occur at time $t_a$, so that $\lambda_{f_i}=1$ for $0\leq t<t_a$, and $\lambda_{f_i}$ switches to a value that lies between $0$ and $1$ for $t>t_a$. It is finally postulated that the higher order effects are such that $f(x)$ is a known vector that can be determined at each instant of time, while $\Phi$ is an unknown vector parameter. Such a dynamic model is often used in flight control problems \cite{lavretsky2013robust}.

Any physical system is subjected to constraints, and actuators in aircraft which are the agencies that are responsible for generating control inputs are not exceptions to this rule either.  As the focus of this paper is the design of a control architecture in the context of anomalies, we explicitly accommodate actuator constraints.  In particular, we assume that the $u$ is assumed to be position / amplitude limited and modelled as
\begin{equation}\label{e:u_i_1}
u_i(t)=u_{\text{max}_i}\text{sat}\left(\frac{u_{c_i}(t)}{u_{\text{max}_i}}\right)=
\left\{\begin{array}{ll}
u_{c_i}(t), & |u_{c_i}(t)|\leq u_{\text{max}_i}\\
u_{\text{max}_i}(t)\text{sgn}\left(u_{c_i}(t)\right), & |u_{c_i}(t)|> u_{\text{max}_i}
\end{array}\right.
\end{equation}
where $u_{\text{max}_i}$ for $i=1,\ldots,m$ are the physical amplitude limits of actuator $i$, and $u_{c_i}(t)$ are the control inputs to be determined by the shared control architecture (SCA). The functions $\text{sat}(\cdot)$ and $\text{sgn}(\cdot)$ denote saturation and sign functions, respectively.

\subsection{Advanced autopilot based on adaptive control}
The control input $u_{c_i}(t)$ will be constructed using an adaptive controller. To specify the adaptive controller, a reference model that specifies the commanded behavior from the plant is constructed and is of the form \cite{narendra2005stable}
\begin{equation}\label{e:reference_model}
\dot{x}_m(t)=A_mx_m(t)+B_mr_0(t),
\end{equation}
where $r_0\in\mathbb{R}^k$ is a reference input, $A_m~(n\times n)$ is a Hurwitz matrix, $x_m\in\mathbb{R}^n$ is the state of the reference model and $(A_m,B_m)$ is controllable. The goal of the adaptive autopilot is then to choose $u_{c_i}(t)$ in (\ref{e:u_i_2}) so that if an error $e$ is defined as
\begin{equation}\label{e:e}
e(t)=x(t)-x_m(t),
\end{equation}
all signals in the adaptive system remain bounded with error $e(t)$ tending to zero asymptotically.

The design of adaptive controllers in the presence of control magnitude constraints is first addressed in \cite{karason1993adaptive}, with guarantees of closed-loop stability through modification of the error used for the adaptive law. The same problem is addressed in \cite{lavretsky2007stable}, using an approach termed \textit{$\mu$-mod adaptive control} where the effect of input saturation is accommodated through the addition of another term in the reference model. Yet another approach based on a closed-loop reference model (CRM) is derived in \cite{lee1997error}, \cite{gibson2013adaptive} in order to improve the transient performance of the adaptive controller. The autopilot we propose in this paper is based on both the $\mu$-mod and CRM approaches. Using the control input modified in (\ref{e:u_i_1}), this controller is compactly summarized as

\begin{equation}\label{e:u_i_2}
u_{c_i}(t)=
\left\{\begin{array}{ll}
u_{\text{ad}_i}(t), & |u_{\text{ad}_i}(t)|\leq u_{\text{max}_i}^{\delta}\\
\frac{1}{1+\mu}\left(u_{\text{ad}_i}(t)+\mu \text{sgn}\left(u_{\text{ad}_i}(t)\right)u_{\text{max}_i}^{\delta}\right), & |u_{c_i}(t)|> u_{\text{max}_i}^{\delta}
\end{array}\right.
\end{equation}
where
\begin{equation}\label{e:adaptive_input}
u_{\text{ad}_i}(t)=K_x^T(t)x(t)+K_r^T(t)r_0(t)+\hat{d}(t)+\hat{\Phi}^T(t)f(x),
\end{equation}
\begin{equation}\label{e:u_max}
u_{\text{max}_i}^{\delta}=(1-\delta)u_{\text{max}_i},\quad 0\leq\delta<1.
\end{equation}

A buffer region in the control input domain $[(1-\delta)u_{\text{max}_i},u_{\text{max}_i}]$ is implied by $(\ref{e:u_i_2})$ and $(\ref{e:u_max})$ and the choice of $\mu$ allows the input to be scaled somewhere in between. The reference model is also modified as
\begin{equation}\label{e:CRM_mod}
\dot{x}_m(t)=A_mx_m(t)+B_m\left(r_0(t)+K_u^T(t)\Delta u_{\text{ad}}(t)\right)-Le(t),
\end{equation}
\begin{equation}\label{e:delta_u_ad}
\Delta u_{\text{ad}_i}(t)=u_{\text{max}_i}\text{sat}\left(\frac{u_{c_i}(t)}{u_{\text{max}_i}}\right)-u_{\text{ad}_i}(t),
\end{equation}
and $L<0$ is a constant or a matrix selected such that $(A_m+L)$ is Hurwitz. Finally, the adaptive parameters are adjusted as
\begin{align}\label{e:adaptive_laws}
\begin{split}
\dot{K}_x(t)&=-\Gamma_xx(t)e^T(t)PB,\\
\dot{K}_r(t)&=-\Gamma_r r_0(t)e^T(t)PB,\\
\dot{\hat{d}}(t)&=-\Gamma_d e^T(t)PB,\\
\dot{\hat{\Phi}}(t)&=-\Gamma_ff(x(t))e^T(t)PB,\\
\dot{K}_u(t)&=\Gamma_u\Delta u_{\text{ad}}e^T(t)PB_m,
\end{split}
\end{align}
where $P=P^T$ is a solution of the Lyapunov equation (for $Q>0$) 
\begin{equation}\label{e:lyap}
A_m^TP+PA_m=-Q,
\end{equation}
with $\Gamma_x=\Gamma_x^T>0$, $\Gamma_r=\Gamma_r^T>0$, $\Gamma_u=\Gamma_u^T>0$.

The stability of the overall adaptive system specified by (\ref{e:plant}), (\ref{e:u_i_1}), (\ref{e:reference_model})-(\ref{e:lyap}) is established in \cite{lavretsky2007stable} when $L=0$. The stability of the adaptive system, when no saturation inputs are present, is also established in \cite{gibson2013adaptive}. A very straightforward combination of the two proofs can be easily carried out to prove that when $L<0$ the adaptive system considered in this paper has globally bounded solutions if the plant in (\ref{e:plant}) is open-loop stable and bounded solutions for an arbitrary plant if all initial conditions and the control parameters in (\ref{e:adaptive_laws}) lie in a compact set. The proof is skipped due to page limitations.

The adaptive autopilot in (\ref{e:u_i_2})-(\ref{e:lyap}) provides the required control input, $u$, in (\ref{e:plant}) as a solution to the underlying problem. The autopilot includes several free parameters including $\mu$ in (\ref{e:u_i_2}), $\delta$ in (\ref{e:u_max}), the reference model parameters $A_m$, $B_m$, $L$ in (\ref{e:CRM_mod}) and the control parameters $K_x(0)$, $K_r(0)$, $K_u(0)$, $\hat{d}(0)$, $\hat{\Phi}(0)$ in (\ref{e:adaptive_laws}). As will be seen in the next section, the parameters $\delta$ and $\mu$ are related to CfM and GCD. 

\subsubsection{Quantification of CfM, GCD and Trade-offs}
The control input $u_{c_{i}}$ in (\ref{e:u_i_2}) is shaped by two parameters $\delta$ and $\mu$, both of which help tune the control input with respect to its specified magnitude limit $u_{\text{max}_i}$. We use these two parameters in quantifying CfM, GCD, and the trade-offs between them as follows.

\begin{itemize}
	\item CfM: As mentioned earlier, qualitatively, CfM corresponds to a system's reserved capacity, which we quantitatively formulate in the current context as the distance between a control input and its saturation limits. In particular, we define CfM as
	\begin{equation}\label{e:unscaled_cfm}
	\text{CfM}=\dfrac{\text{CfM}^{R}}{\text{CfM}_{d}}, 
	\end{equation}
	where
	\begin{align}\label{e:unscaled_rms_cfm}
	\begin{split}
	\text{CfM}^R&=\text{rms}\left(\min_i(c_i(t))\right)\Big|_{t_a}^T,\\
	c_i(t) & =1 - \dfrac{|u_i(t)|}{u_{\text{max}_i}},
	\end{split}
	\end{align}
	
	$c_{i}(t)$ is the normalized CfM variation of actuator $i$, $\text{CfM}^{\text{R}}$ is the root mean squared CfM variation and CfM$_d$, which denotes the desired CfM is chosen as 
	\begin{equation}\label{e:cfm_desired}
	\text{CfM}_{d}=\max_i\left(\delta u_{\text{max}_i}\right).
	\end{equation}
	
	In the above equations, $\min$ and $\max$ are the minimum and maximum operators over the $i^{th}$ index, $\text{rms}$ is the root mean square operator and $t_a$ and $T$ refer to the time of anomaly and final simulation time, respectively. From (\ref{e:unscaled_rms_cfm}), we note that ($i$) $\text{CfM}^{R}$ has a maximum value $u_{\text{max}}$ for the trivial case when all $u_i(t)=0$, ($ii$) a value close to $\delta u_{\text{max}}$ if the control inputs approach the buffer region, and ($iii$) zero if $u_i(t)$ hits the saturation limit $u_{\text{max}}$. Since $\text{CfM}_d=\delta u_{\text{max}}$, it follows that CfM, the corresponding normalized value, is greater than unity when the control inputs are small and far away from saturation, unity as they approach the buffer region, and zero when fully saturated. The buffer region $[(1-\delta)u_{\text{max}_{i}}, \enspace u_{\text{max}_{i}}]$ that $|u_i(t)|$ belongs to can be mapped into the CfM domain as a virtual buffer $[0,\enspace \delta]$ by substituting the upper and lower limit $u_{\text{max}}$ and $u_{\text{max}}^{\delta}$ into $|u_{i}(t)|$ in (\ref{e:unscaled_rms_cfm}).
	
	\item GCD: As mentioned earlier, the reference model represents the commanded behavior from the plant being controlled.  In order to reflect the fact that the actual output may be compromised if the input is constrained, we add a term that depends on $\Delta u_{\text{ad}}(t)$ in (\ref{e:delta_u_ad}) to become nonzero whenever the control input saturates, that is, when the control input approaches the saturation limit, $\Delta u_{\text{ad}_i}$ becomes nonzero, thereby suitably allowing a graceful degradation of $x_m$ from its nominal choice as in (\ref{e:CRM_mod}). We denote this degradation as GCD and quantify it as:
	\begin{equation}\label{e:gcd_var}
	\text{GCD}=\dfrac{\text{rms}(x_m(t)-r_0(t))}{\text{rms}(r_0(t))},\quad t\in T_0,
	\end{equation}
	
	where $T_0$ denotes the interval of interest. It should be noted that once $\mu$ is specified, the adaptive controller automatically scales the input into the reference model through $\Delta u_{\text{ad}}$ and $K_u$, in a way so that $e(t)$ remains small and the closed-loop system has bounded solutions.\\
	
	\item $\mu$: The intent behind the introduction of the parameter $\mu$ in (\ref{e:u_i_2}) is to regulate the control input and move it away from saturation when needed. For example, if $|u_{ad_i}(t)|>u_{\text{max}_i}^{\delta}$, the extreme case of $\mu=0$ will simply set $u_{c_i}=u_{\text{ad}}$, thereby removing the effect of the virtual limit imposed in (\ref{e:u_max}). As $\mu$ increases, the control input would decrease in magnitude and move towards the virtual saturation limit $u_{\text{max}_i}^{\delta}$, that is, once the buffer $\delta$ is determined, $\mu$ controls $u_i(t)$ within the buffer region $[(1-\delta)u_{\text{max}_i},u_{\text{max}_i}]$, bringing it closer to the lower limit with increasing $\mu$. In other words, as $\mu$ increases, CfM increases as well in the buffer region.\\
	
	It is easy to see from (\ref{e:CRM_mod}) and (\ref{e:delta_u_ad}) that similar to CfM, as $\mu$ increases, GCD increases as well. This is due to the fact that an increase in $\mu$ increases $\Delta u_{\text{ad}_i}(t)$ which, in turn, increases the GCD. While a larger CfM improves the responsiveness of the system to future anomalies, a lower bound on the reference command is necessary to finish the mission within practical constraints. In other words, $\mu$ needs to be chosen so that GCD remains above a lower limit while maintaining a large CfM. We relegate the task of selecting the appropriate $\mu$ to the human pilot.
\end{itemize}

\subsubsection{Choice of the Reference Model Parameters}

In addition to $\mu$ and $\delta$, the adaptive controller in \eqref{e:u_i_1}, \eqref{e:reference_model}-\eqref{e:lyap} requires the reference model parameters $A_m$, $B_m$, $L$ and the control parameters $K_x(0)$, $K_r(0)$, and $K_u(0)$ at time $t=0$. If no anomalies are present, then $\Lambda_{\text{nom}}=\Lambda_f=I$ which implies that $A_m$ and $B_m$ as well as the control parameters can be chosen as
\begin{align}\label{e:matching}
\begin{split}
A_m&=A+BK_x^T(0),\\
K_r^T(0)&=-(A_m^{-1}B)^{-1},\\
B_m&=BK_r^T(0),\\
K_u^T(0)&=-A_m^{-1}B,
\end{split}
\end{align}
where $K_x(0)$ is computed using a linear–quadratic regulator (LQR) method and the nominal plant parameters $(A, B)$ \cite{bryson1996optimal} and $K_r(0)$  in (\ref{e:matching}) is selected to provide unity low frequency DC gain for the closed-loop system. When anomalies occur, $\Lambda_f \neq I$, at time $t=t_a$ and supposing that an estimate $\hat{\Lambda}_f$ is available, a similar choice as in (\ref{e:matching}) can be carried out using the plant parameters $(A,B\hat{\Lambda}_f)$ and the relations
\begin{align}\label{e:matching2}
\begin{split}
A_m&=A+B\hat{\Lambda}_fK_x^T(t_a),\\
K_r^T(t_a)&=-(A_m^{-1}B\hat{\Lambda}_f(t_a))^{-1},\\
B_m&=B\hat{\Lambda}_fK_r^T(t_a),\\
K_u^T(t_a)&=-A_m^{-1}B\hat{\Lambda}_f(t_a),
\end{split}
\end{align}
with the adaptive controller specified using (\ref{e:u_i_1})-(\ref{e:lyap}) for all $t\geq t_a$. Finally, $L$ is chosen as in \cite{gibson2013adaptive} and lower parameters $\hat{d}(0)$, $\hat{\Phi}(0)$ are chosen arbitrarily. Similar to $\mu$, we relegate the task of assessing the estimate $\hat{\Lambda}_{f_{p}}$ to the human pilot as well.

\subsection{Autopilot based on proportional derivative control}
To investigate shared control architectures, another autopilot that is employed in the closed loop system is the proportional derivative (PD) controller. Assuming a single control input \cite{farjadian2016resilient}, the goal is to control the dynamics \cite{hess2009modeling}
\begin{equation} \label{e:hess1} 
Y_p(s)=\frac{1}{s(s+a)}, 
\end{equation}
which represents the aircraft transfer function between the input $u$ and an output $M(t)$, and can be assumed to be a simplified version of the dynamics in (1). The input $u$ is subjected to the same magnitude and rate constraints as in \eqref{e:u_i_1}. Based on \eqref{e:hess1}, a PD controller can be chosen as

where the same actuator position and rate constraints explained in  \eqref{e:u_i_1} are assumed. Considering the transfer function \eqref{e:hess1} between the input $u$ and the output $M(t)$, the PD controller can be chosen as 
\begin{equation} \label{pdcontrol}
u(t)= K_p (M(t)-M_{\text{cmd}}(t))+ K_r (\dot M(t)),
\end{equation} 
where $M_{\text{cmd}}$ is the desired command signal that $M$ is required to follow. Given the second-order structure of the dynamics, it can be shown that suitable gains $K_p$ and $K_r$ can be determined so that the closed-loop system is stable and for command inputs at low frequencies, a satisfactory tracking performance can be obtained.

It is noted that during the experimental validation studies certain anomalies are introduced to \eqref{e:hess1} in the form of unmodeled dynamics and time delays. Therefore, according to the crossover model \cite{mcruer1974mathematical}, the human pilot has to adapt themselves to demonstrate different compensation characteristics, such as pure gain, lead or lag, based on the type of the anomaly. This creates a challenging scenario for the shared control architecture.

\section{Human Pilot}

In this section, we discuss mathematical models of human pilot decision making on the basis of absence and presence of flight anomalies. A great deal of research has been conducted on mathematical human pilot modeling assuming that no failure in the aircraft or no severe disturbances in the environment are present. Since decision making will differ significantly whether the aircraft is under nominal operation or subjected to severe anomalies, the corresponding models are entirely different as well and are discussed separately in what follows.

\subsection{Pilot Models in the Absence of Anomaly}

In the absence of anomalous event(s), the mathematical models of human pilot control behavior can fundamentally be assorted according to control-theoretic, physiological and, more recently, machine learning methods \cite{lone2014review}, \cite{xu2017review}. One of the most renowned control-theoretic method in the modelling of human pilot, namely, \textit{the crossover model}, is presented in \cite{mcruer1974mathematical} as an assembly of the pilot and the controlled vehicle, for single loop control systems. The open loop transfer function for the crossover model is
\begin{equation}\label{crossovermodel}
Y_{h}(j\omega)Y_{p}(j\omega)=\dfrac{\omega_{c}e^{-\tau_{e}j \omega}}{j \omega},
\end{equation}
where $Y_{h}(j\omega)$ is a transfer function of the human pilot, $Y_{p}(j\omega)$ is a transfer function of the aircraft, $\omega_{c}$ is the crossover frequency and $\tau_{e}$ is the effective time delay pertinent to the system delays and human pilot lags. The crossover model is applicable for a range of frequencies around the crossover frequency, $\omega_{c}$. When a ``remnant'' signal is introduced to $Y_h(j\omega)$ to account for the nonlinear effects of the pilot-vehicle system, the model is called a quasi-linear model \cite{mcruer1967review}.

Other sophisticated quasi-linear models can be found in the literature as \textit{the extended crossover model} \cite{mcruer1967review}, which works especially for conditionally stable systems, that is, when a pilot attempts to stabilize an unstable transfer function of the controlled element, $Y_{p}(j\omega)$, and \textit{the precision model} \cite{mcruer1967review}, which treats a wider frequency region than the crossover model. The single loop control tasks are covered by quasi-linear models.  They can be extended to multiloop control tasks by the introduction of \textit{the optimal control model} \cite{wierenga1969evaluation}, \cite{kleinman1970optimal}. This model enables the controlled elements to be given in state space representation provided that a well-trained pilot controls the aircraft in an optimal fashion.

Another approach in modelling the human pilot is employing the information of sensory dynamics relevant to humans to extract the effect of motion, proprioceptive, vestibular and visual cues on the control effort. An example of this is \textit{the descriptive model} \cite{hosman1998pilot} in which a series of experiments are conducted to distinguish the influence of vestibular and visual stimuli from the control behavior. Another example can be given as \textit{the revised structural model} \cite{hess1997unified} where the human pilot is modeled as the unification of proprioceptive, vestibular and visual feedback paths. It is hypothesized that such cues help alleviate the compensatory control action taken by the human pilot.

It is noted that due to their physical limitations, models such as in \eqref{crossovermodel} are valid for operation  in a predefined boundary or envelope where the environmental factors are steady and stable. In the case of an anomaly, they may not perform as expected \cite{woods2000learning}.

\subsection{Pilot Models in the Presence of Anomaly} 

When extreme events and failures occur, human pilots are known to adapt themselves to changing environmental conditions, which overstep the boundaries of automated systems. Since the hallmark of any autonomous system is its ability to self-govern even under emergency conditions, modeling of the pilot decision-making upon occurrence of an anomaly is indispensable, and examples are present in the literature \cite{hess2009modeling, hess2014model, hess2016modeling, farjadian2016resilient, farjadian2017bumpless, farjadian2018resilient, thomsen2019shared, tohidi2019adaptive}. In this paper, we assume that the anomalies can be modeled either as an abrupt change in the vehicle dynamics \cite{hess2009modeling, hess2014model, hess2016modeling,thomsen2019shared, farjadian2016resilient, tohidi2019adaptive}, or a loss of control effectiveness in the control input \cite{farjadian2017bumpless, farjadian2018resilient}.  In either case, the objective is the modeling of the decision making of the pilot so as to elicit a resilient performance from the aircraft and recover rapidly from the impact of the anomaly. Since the focus of this article is shared control, among the pilot models that are developed for anomaly response, we exploit the ones that explains the pilot behavior in relation to the autopilot. These models are developed using the Capacity for Maneuver (CfM) concept, the details of which are explained in the following sections.

% Original Version (to be deleted)
% When extreme events and failures occur, human pilots are known to adapt themselves to changing environmental conditions, which overstep the boundaries of automated systems. Since the hallmark of any autonomous system its ability to self-govern even under emergency conditions, modeling of the pilot decision-making upon occurrence of an anomaly is indispensable, and examples are present in the literature \cite{hess2009modeling, hess2014model, hess2016modeling, farjadian2018resilient, thomsen2019shared}. In this paper, we assume that the anomalies can be modeled either as an abrupt change in the vehicle dynamics \cite{hess2009modeling, hess2014model, hess2016modeling,thomsen2019shared}, or a loss of control effectiveness in the control input \cite{farjadian2018resilient}.  In either case, the objective is the modeling of the decision making of the pilot so as to elicit a resilient performance from the aircraft and recover rapidly from the impact of the anomaly.

\subsubsection{Pilot Models Based on Capacity for Maneuver (CfM)}

A recent method in modelling of human pilot under anomalous events utilizes the Capacity for Maneuver (CfM) concept \cite{farjadian2016resilient}, \cite{farjadian2017bumpless}, \cite{farjadian2018resilient}. As discussed in the ``Autopilot'' section, CfM refers to the remaining range of the actuators before saturation, which quantifies the available maneuvering capacity of the vehicle. It is hypothesized that surveillance and regulation of a system’s available capacity to respond to all events help maintain the resiliency of a system, which is a necessary merit to recover from unexpected and abrupt failures or disturbances \cite{woods2018theory}. We propose two different types of pilot models, both of which use CfM, but in different ways.

%\textcolor{violet}{I added the word file as promised but paraphrased it a little bit and also added the definitions of $G_{1}(s)$ etc.}

\begin{itemize}
	\item \textit{Perception trigger}: Here, the pilot model is assumed to assess the CfM and implicitly compute a perception gain based on the CfM. The quantification of this gain, $K_t$, is predicated on CfM$^R$ with the definition as in (\ref{e:unscaled_rms_cfm}). The perception trigger is associated to the gain $K_t$, which is implicitly computed as in \cite{farjadian2016resilient}. The perception algorithm for the pilot is  
	\begin{equation}\label{e:perception_trigger}
	K_t=
	\left\{\begin{array}{ll}
	0, & |F_{0}|< 1\\
	1, & |F_{0}|\geq 1
	\end{array}\right.
	\end{equation}
	where
	\begin{equation}\label{perception_variable1}
	F_{0}=G_{1}(s)[F(t)],
	\end{equation}
	and
	\begin{equation}\label{perception_variable2}
	F(t)=\dfrac{\dfrac{d}{dt}(\text{CfM}^{R})-\mu_{p}}{3\sigma_{p}}.
	\end{equation}
	$G_{1}(s)$ is a second order filter introduced as a smoothing and lagging operator into human perception algorithm, $F(t)$ is the perception variable, $\mu_{p}$ is the average of $\frac{d}{dt}(\text{CfM}^R)$ and $\sigma_{p}$ is the standard deviation of $\frac{d}{dt}(\text{CfM}^R)$, both of which are measured over a nominal flight simulation. The computation of these statistical parameters is further elaborated in subsection ``Validation of Shared Control Architecture 1". The hypothesis here is that the human pilot has such a perception trigger, $K_t$, and when this trigger, $K_t=1$, the pilot takes over control from the autopilot. In Section ``Validation of Shared Control Architecure 1", we validate this perception model.\\

	\item \textit{CfM-GCD Tradeoff}: Here, the pilot is assumed to implicitly assess the available (normalized) CfM when an anomaly occurs, and decide on the amount of GCD that is allowable so as to let the CfM become comparable to the $\text{CfM}_d$. In other words, we assume that the pilot is capable of assessing the parameter $\mu$ and input this value to the autopilot following the occurrence of an anomaly. In Section "Validation of Shared Control Architecture 2", we validate this assessment.
\end{itemize}

\section{Shared Control}
The focus of this paper is on a shared control architecture that combines the decision making of a pilot and autopilot in flight control. The architecture is invoked under alert conditions,  with triggers in place that specify when the decision-making is transferred from one authority to another. The specific alert conditions that we focus on in this paper corresponds to physical anomalies that compromise the actuator effectiveness. The typical roles of the autopilot and the pilot in a flight control problem were described in Sections ``Autopilot" and ``Human Pilot" respectively. In Section ``Autopilot", autopilots based on PD control and adaptive control were described, with former designed to ensure satisfactory command following under nominal conditions, and the latter to accommodate parametric uncertainties including loss of control effectiveness in the actuators. In Section ``Pilot Models in the Presence of Anomaly", two different models of decision making in pilots were proposed, both based on the monitoring of CfM of the actuators. In this section, we propose two different shared control architectures, using the models of the autopilots and pilots described in the previous sections.

\subsection{Shared Control Architecture 1: A pilot with a CfM based perception and a fixed-gain autopilot}

\begin{figure}[htb]
	\centering
	\includegraphics[scale=0.96]{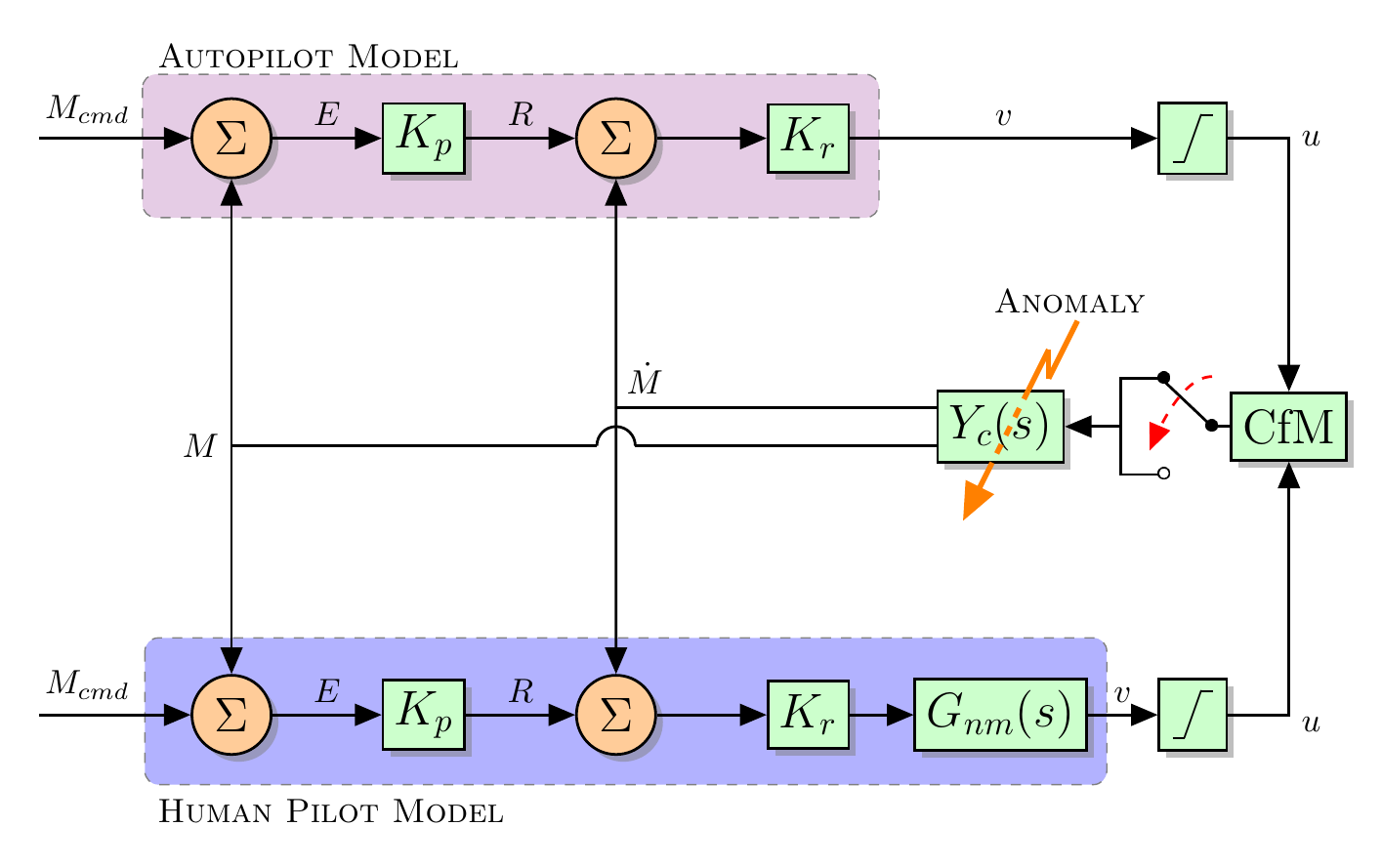}
	
	\caption{Block diagram of the shared control architecture 1 (adapted from \cite{farjadian2016resilient}). The autopilot model consists of a fixed gain controller, whereas the human pilot model comprehends a perception part based on Capacity for Maneuver (CfM) concept and an adaptation part governed by empirical adaptive laws \cite{hess2016modeling}. The block $G_{\text{nm}}(s)$ refers to a simple model of neuromuscular dynamics of a pilot \cite{hess2006simplified}. The neuromuscular transfer function, $G_{\text{nm}}(s)$, which corresponds to control input formed by an arm or leg is given as $G_{\text{nm}}(s)=\frac{100}{s^2+14.14s+100}$. When an anomaly occurs, the plant dynamics, $Y_{\text{p}}(s)$, undergoes an abrupt change by rendering the autopilot insufficient for the rest of the control. At this stage, the occurrence of an anomaly is captured by the CfM such that the control is handed over to the human pilot model for a resilient flight control.}
	\label{fig:shared_combined}
\end{figure}

The first shared control architecture can be summarized as a sequence \{autopilot runs, anomaly occurs, pilot takes over\}. That is, it is assumed that an autopilot  based on PD control as in (\ref{pdcontrol}) is in place, ensuring a satisfactory command tracking under nominal conditions. The human pilot is assumed to consist of a perception component and an adaptation component. The perception component consists of monitoring CfM$^R$, through which a perception trigger $F_0$ is calculated using (\ref{perception_variable1}) and (\ref{perception_variable2}). The adaptation component consists of monitoring the control gain in (\ref{e:perception_trigger}), and taking over control of the aircraft when $K_t=1$. The details of this shared controller and its evaluation using a numerical simulation study can be found in \cite{farjadian2016resilient}. Figure $\ref{fig:shared_combined}$ illustrates the schematic of SCA1.

\subsection{Shared Control Architecture 2: A pilot with a CfM-based decision making and an advanced adaptive autopilot }

\begin{figure}[htb]
	\centering	\includegraphics[scale=1.10]{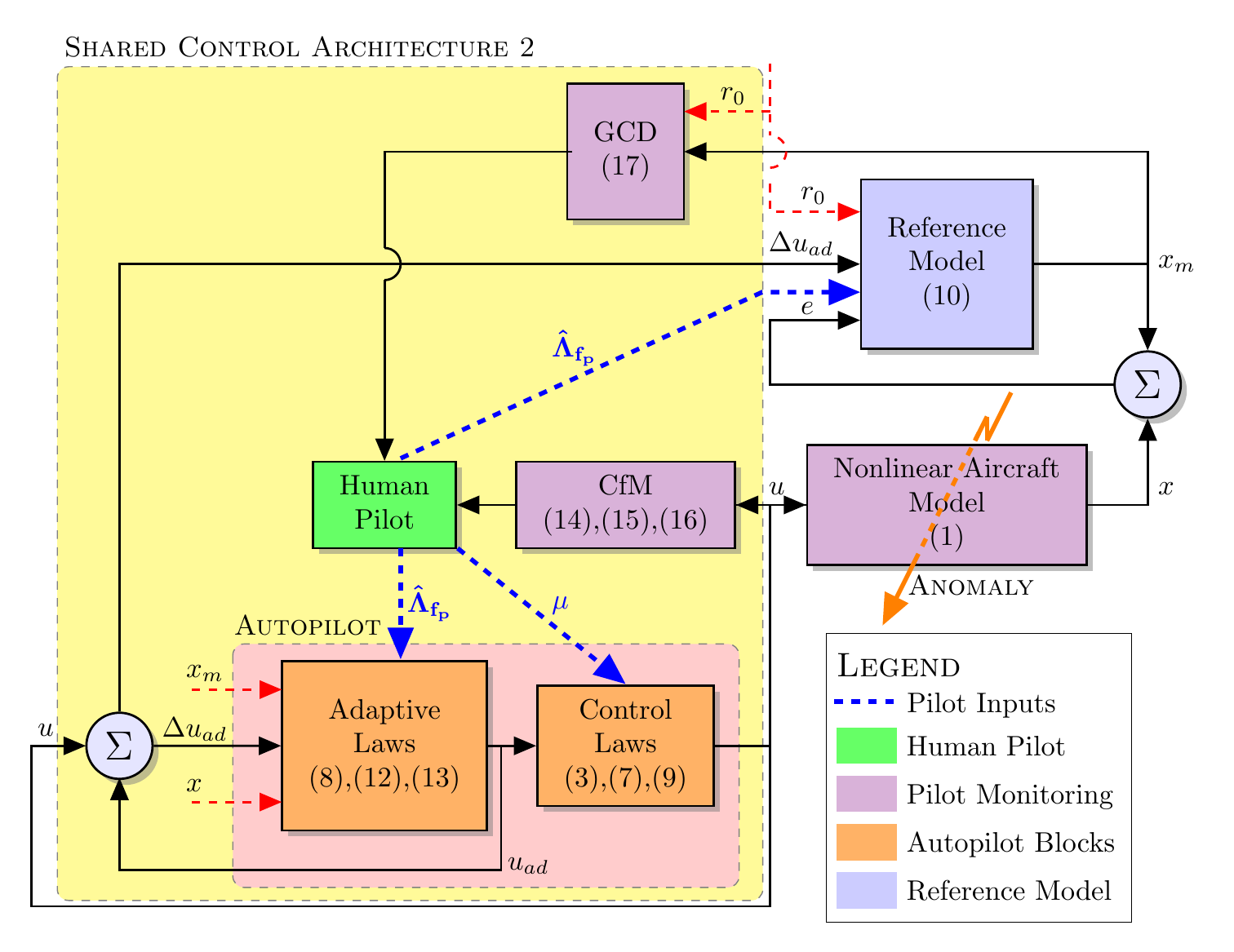}
	
	\caption{Block diagram of the proposed shared control architecture (SCA) 2. The human pilot undertakes a supervisory role by providing the key parameters $\mu$ and $\hat \Lambda{_{f_p}}$ to the adaptive autopilot. The blocks are expressed in different colors based on their functions in the proposed SCA2. The numbers in parentheses in each block correspond to the related equations.}
	\label{fig:SCA2_Detailed}
\end{figure}

In this shared control architecture, the role of the pilot is a supervisory one while the autopilot takes on an increased and more complex role. The pilot is assumed to monitor the CfM of the resident actuators in the aircraft following an anomaly. In an effort to allow the CfM to stay close to the CfM$_d$ in (\ref{e:cfm_desired}), the command is allowed to be degraded; the pilot then determines a parameter $\mu$ which directly scales the control effort through (\ref{e:u_i_2}) and indirectly scales the command signal through (\ref{e:CRM_mod}) and (\ref{e:delta_u_ad}). Once $\mu$ is specified by the pilot, then the adaptive autopilot continues to supply the control input using (\ref{e:u_i_2})-(\ref{e:lyap}). If the pilot has high situational awareness, he/she provides $\Lambda_{f_{p}}$ as well, which is an estimate of the severity of the anomaly. The details of this shared controller and its evaluation using a numerical simulation study can be found in \cite{farjadian2017bumpless}, \cite{farjadian2018resilient}. Figure \ref{fig:SCA2_Detailed} shows the schematic of SCA2. 

%\subsection{Shared Control Architecture 3: A pilot with a latency-based decision making and an advanced adaptive autopilot } The structure of SCA 3 is fairly similar to that of SCA2 except in the specific parameters that the pilot chooses to monitor as well as transmits to the autopilot. The autopilot is based on an adaptive controller as well, though different from that in (\ref{e:u_i_2})-(\ref{e:lyap}). Rather than utilization of CfM to monitor the alert status of the aircraft, the intrinsic latency in the aircraft response to control inputs is proposed to be monitored by the pilot. This information is then used by the pilot to map into a parameter $n^*$ that is transmitted to the autopilot. The autopilot in turn uses this parameter to switch to a specific adaptive controller whose order, number of adjustable parameters, and overall structure are suitably chosen. The details of this shared controller and its evaluation using a numerical simulation study can be found in \cite{thomsen2019shared}. 

%We restrict our attention to the validation of SCA1 and SCA2 in the following section while that of SCA 3 is relegated to future work.

\section{Validation with Human In the Loop Simulations}

The two SCAs we presented in the earlier section correspond to a human pilot working together with an autopilot in order to combat the effects of an anomaly. In SCA1, we hypothesize a human pilot model in the form of \eqref{e:perception_trigger}-\eqref{perception_variable2}, and propose that when an anomaly occurs, the pilot monitors the perception trigger $K_t$ and when it becomes unity, takes over the helm from the autopilot and ensures a safe performance. In SCA2,  an advanced adaptive-control based autopilot is assumed to be in effect, and the pilot is hypothesized to monitor the CfM and input a key parameter $\mu$ to the autopilot. The autopilot in turn leads the closed-loop system to a safe performance with this key input. As nominal (anomaly-free) plant dynamics,  the simpler model  introduced in \eqref{e:hess1} is used for SCA1 validation, and a nonlinear F-16 dynamics \cite{stevens2015aircraft}, \cite{nguyen1979simulator} is used for SCA2 validation. The goal in this section is to validate the two hypotheses of the human-pilot actions. Despite the presence of obvious common elements to the two SCAs, of a human pilot, an autopilot, and a shared controller that combines their decision making, all of the details differ significantly. We therefore describe the validation of these two hypotheses separately in what follows. In each case, we present the validation in the following order: the experimental setup, the type of anomaly, the experimental procedure, details of the human subjects, the pilot-model parameters, results and observations.

Although the experimental procedures used for SCA1 validation and SCA2 validation differ, and therefore explained in separate subsections, they use similar principles. To minimize repetition, main tasks can be summarized as follows: The procedure comprises three main phases, namely, \textit{Pilot Briefing}, \textit{Preparation Tests} and \textit{Performance Tests}. In \textit{Pilot Briefing}, the overall aim of the experiment and the experimental setup are introduced. In \textit{Preparation Tests}, the subjects are encouraged to gain practice with the joystick controls. In the last phase, \textit{Performance Tests}, the subjects are expected to conduct the experiments only once in order not to affect the reliability due to learning. Anomaly introduction times are randomized to prevent predictability. 

It is noted that during the validation studies, SCA1 and SCA2 are not compared with each other. They are separately validated with different human-in-the-loop simulation settings. However, comparison of SCA1 and SCA2 with each-other can be pursued as a future research direction.
\begin{figure}[htb]
	\centering
	\includegraphics[scale=0.15]{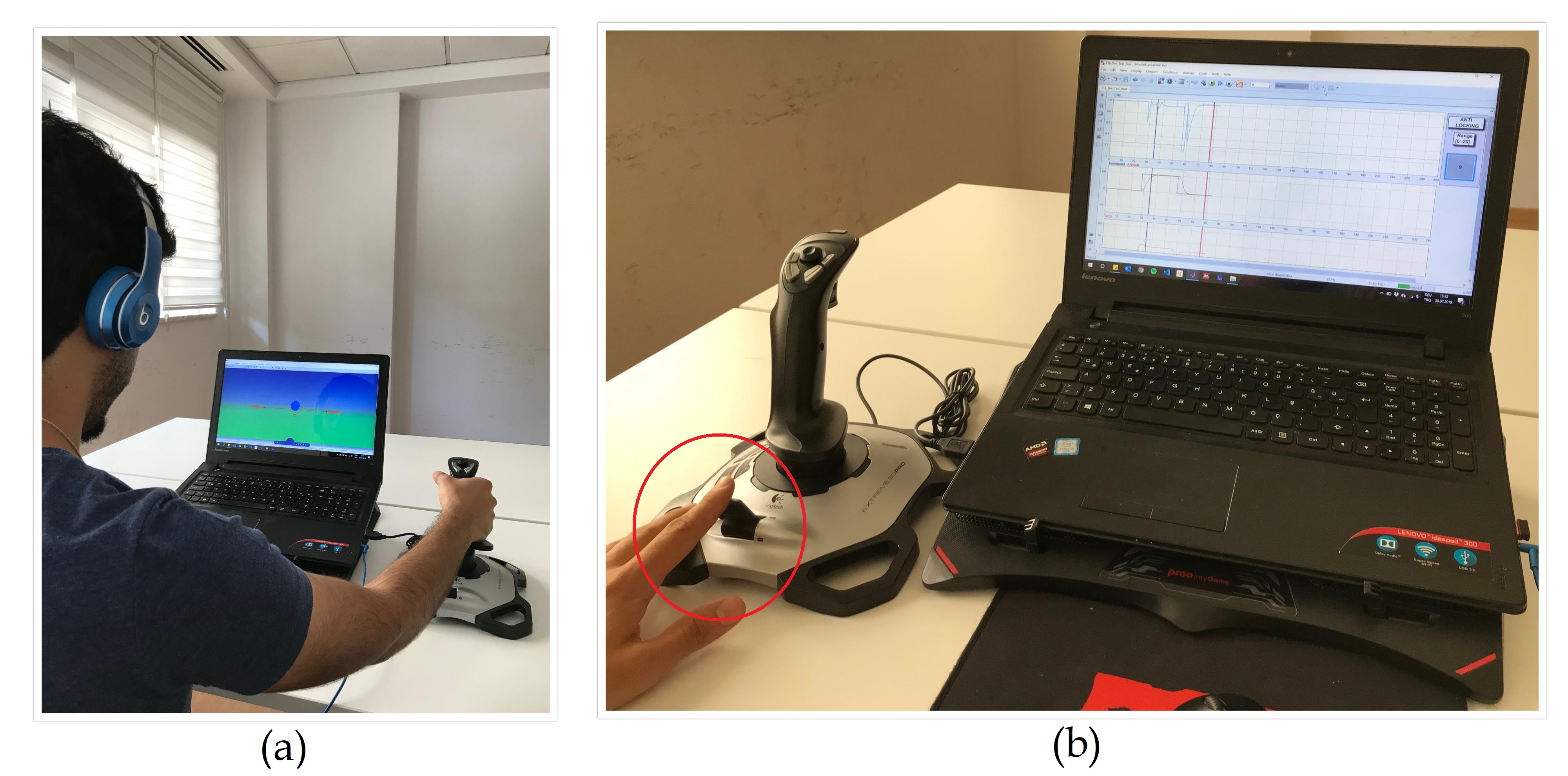}
	\caption{Experimental setups for shared control architecture (SCA) 1 (a) and 2 (b). SCA1 experiment consists of a pilot screen (see Figure \ref{fig:pilot_screen_approach1}) and a commercially available pilot joystick, whose pitch input is used. SCA2 experiment consists of a different pilot screen (see Figure \ref{fig:simscreen}) and the joystick. In SCA2, subjects use the joystick lever to provide input.}
	\label{fig:collage_experiments}
\end{figure}
\subsection{Validation of Shared  Control  Architecture 1}

\subsubsection{Experimental Setup}
\begin{figure}[htb]
	\centering
	\includegraphics[scale=0.65]{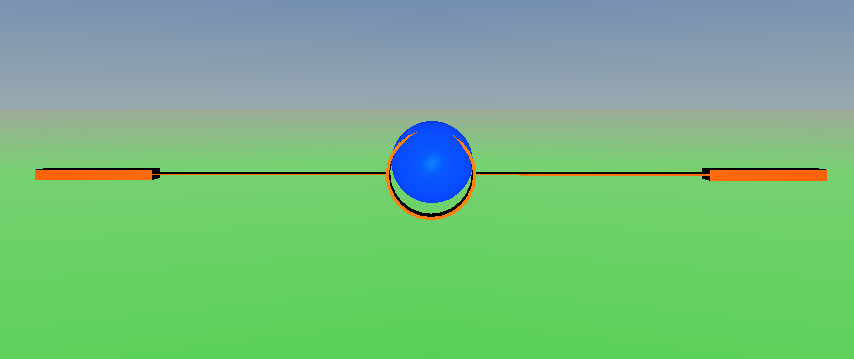}
	
	\caption{Flight screen interface for Shared Control Architecture 1. The blue sphere represents the aircraft nose and is controlled in the vertical direction via joystick movements. The orange line with a circle in the middle, which also moves in the vertical direction, represents the reference command to be followed. The control objective is to keep the blue sphere inside the orange circle. At the beginning of the experiments, the aircraft is controlled by the autopilot. When an anomaly occurs, the subjects are warned to take over the control, via a sound signal. The time of initiating this signal is determined using different alert times.}
	\label{fig:pilot_screen_approach1}
\end{figure}

The experimental setup consists of a pilot screen and the commercially available pilot joystick Logitech Extreme 3D Pro (see Figure~\ref{fig:collage_experiments}) with the goal of performing a desktop, human-in-the-loop, simulation. The flight screen interface for SCA1 can be observed in Figure~{\ref{fig:collage_experiments}~-~a} and separately illustrated in Figure $\ref{fig:pilot_screen_approach1}$. The orange line with a circle in the middle shows the reference to be followed by the pilot. This line is moved up and down according to the desired reference command, $M_{\text{cmd}}$ (see Figure $\ref{fig:shared_combined}$). The blue sphere in Figure $\ref{fig:pilot_screen_approach1}$ represents the nose tip of the aircraft and is driven by the joystick inputs. When the subject moves the joystick, s/he provides the control input, $v$, which goes through the aircraft model and produces the movement of the blue sphere. The control objective is to keep the blue sphere inside the orange circle, which translates into tracking the reference command. Similarly, the flight screen for SCA2 is shown in Figure~{\ref{fig:collage_experiments}~-~b} and separately illustrated in Figure~\ref{fig:simscreen}. The details on this screen are provided in the ``Validation of the Shared Control Architecture 2'' subsection. 

\subsubsection{Anomaly}
The anomaly is modeled by a sudden change in vehicle dynamics, from $Y_p^{\text{before}}(s)$ to $Y_p^{\text{after}}(s)$ as in \cite{hess2016modeling}, and illustrated in Figure \ref{fig:shared_combined}. Two different flight scenarios, $S_{\text{harsh}}$, and $S_{\text{mild}}$, are investigated, which correspond to a harsh and a mild anomaly, respectively. In the harsh anomaly, it is assumed that 
\begin{equation}\label{e:harsh_transfer}
Y_{p,h}^{\text{before}}(s) = \frac{1}{s(s+10)}, \quad Y_{p,h}^{\text{after}}(s) = \frac{e^{-0.2s}}{s(s+5)(s+10)},  
\end{equation}
\noindent whereas in the mild anomaly, it is assumed that
\begin{equation}\label{e:mild_transfer}
Y_{p,m}^{\text{before}}(s) = \frac{1}{s(s+7)}, \quad Y_{p,m}^{\text{after}}(s) = \frac{e^{-0.18s}}{s(s+7)(s+9)}.
\end{equation}
\noindent The specific numerical values of the parameters in \eqref{e:harsh_transfer} and \eqref{e:mild_transfer} are chosen so that the pilot action has a distinct effect in the two cases based on their response time. More details of these choices are provided in the following section.

The anomaly is introduced at a certain instant of time, $t_{a}$, in the experiment. The anomaly alert is conveyed as sound signal at $t_{s}$, following which, the pilots take over control at a time $t_{\text{TRT}}$, after a certain reaction time $t_{\text{RT}}$. Denoting $\Delta T = t_s-t_a$ as the alert time, the total elapsed time from the onset of anomaly to the instant of hitting the joystick button is defined as
\begin{equation}\label{e:trt}
t_{\text{TRT}} = t_{\text{RT}}+ \Delta T.
\end{equation}
%
% ***** Experimental Procedure *****
\subsubsection{Experimental Procedure}
\begin{figure}[htb]
	\centering
	\includegraphics[scale=1.05]{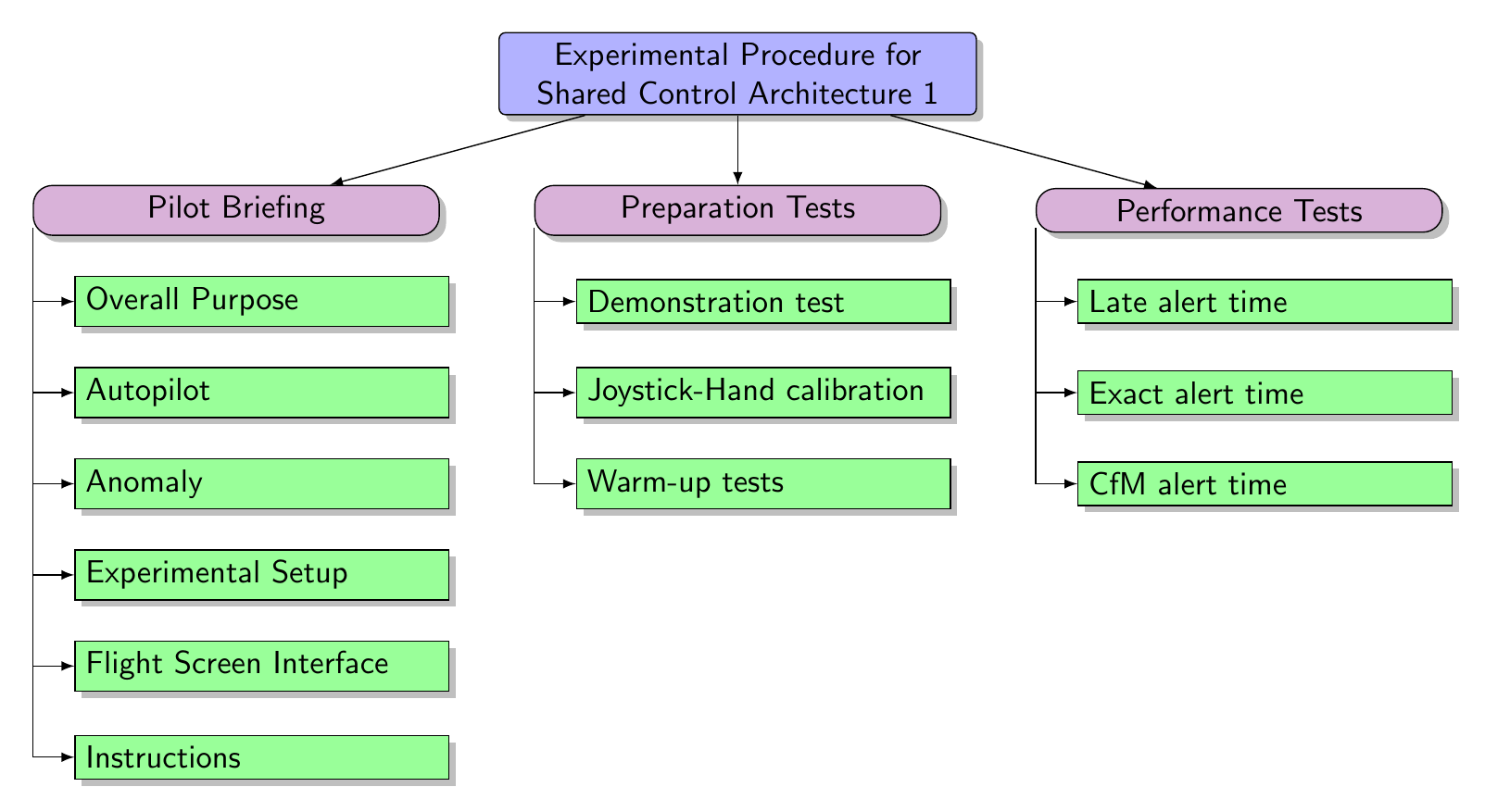}
	\caption{Experimental procedure breakdown for the Shared Control Architecture 1.  Three main tasks constituting the procedure are seen. In \textit{Pilot Briefing}, the subjects read the pilot briefing, review it with the experiment designer and have a question and answer session. In \textit{Preparation Tests}, the subjects get familiarized with the test via demonstration runs and warm-up tests. Finally, in \textit{Performance Tests}, the real tests are conducted using three different alert times.}
	\label{fig:Breakdown_Structure_SCA1}
\end{figure}

The experimental procedure consists of three main parts, which are \textit{Pilot Briefing}, \textit{Preparation Tests} and \textit{Performance Tests} (see Figure $\ref{fig:Breakdown_Structure_SCA1}$).

% Sidebar Comment:
% In \textit{Pilot Briefing}, the subjects read the pilot briefing, the details of which are given in Sidebar "\nameref{sidebar-HITLS1}", review it with...

The first part of the procedure is the \textit{Pilot Briefing}, where the subjects are required to read a pilot briefing to have a clear understanding of the experiment. The briefing consists of six main sections, namely, \textit{Overall Purpose} , \textit{Autopilot}, \textit{Anomaly}, \textit{Experimental Setup}, \textit{Flight Screen} and \textit{Instructions}. In these sections, the main concepts and experimental hardware such as the pilot screen and the joystick lever are introduced to the subject. A scaled toy aircraft prototype is used to help the subjects visualize the concepts covered in the briefing.

% Sidebar Comment:
% A scaled toy aircraft prototype is used to help the subjects visualize the concepts covered in the briefing. All the details regarding the pilot briefing can be found in Sidebar "\nameref{sidebar-HITLS1}".

The second part is the \textit{Preparation Tests}, in which the subjects are introduced to a demonstration test conducted by the experiment designer to familiarize the subjects to the setup. In this test, the subjects observe the experiment designer follow a reference command using the joystick (See Figure \ref{fig:collage_experiments}(a)). They also watch the designer to respond to control switching alert sounds by taking over the control via the joystick. Following these demonstration tests, the subjects are requested to perform the experiment themselves. To complete this part, three preparation tests, each with a duration of $90$ seconds are conducted. At the end of each test, the root mean squared error, $\text{e}_{\text{rms}}$, of the subjects is calculated as
\begin{equation}\label{rms-error}
e_{\text{rms}}= \sqrt{\frac{1}{T_p} \int_{t_{a}}^{T_p} e(\tau)^{2}d\tau},
\end{equation}
where
\begin{equation}\label{error}
e(t) = M_{\text{cmd}}(t)-M(t),
\end{equation}
and $T_p=90$s. It is expected that the $\text{e}_{\text{rms}}$ in each trial decreases as a sign of learning.

The third is the \textit{Performance Tests}, each with a duration of $180$ seconds, which aim at testing the performance of the proposed SCA, in terms of tracking error $e_{\text{rms}}$, CfM and a bumpless transfer metric, $\rho$, which is calculated by taking the difference of $\text{e}_{\text{rms}}$ values that are obtained using the 10 second intervals before and after the anomaly. This calculation is performed as
\begin{equation}\label{rho-bumpless}
\rho= \sqrt{\frac{1}{t_a+10} \int_{t_{a}}^{t_a+10} e(\tau)^{2}d\tau} -  \sqrt{\frac{1}{t_a} \int_{t_a-10}^{t_{a}} e(\tau)^{2}d\tau}.
\end{equation}

% ***** Experimental Procedure *****

\subsubsection{Details of the Human Subjects}
The experiment with the harsh anomaly was conducted by 15 subjects (including 1 flight pilot), whereas the one with the a mild anomaly was conducted by 3 subjects (including 1 flight pilot). All subjects were over 18 years old and 4 of the subjects were left-handed, yet this did not bring about any problems, since an ambidextrous joystick was utilized. The experiment was approved by the Bilkent University Ethics Committee and an informed consent was taken from each subject before conducting the experiment. Some statistical data pertaining to the subjects are given in Table \ref{tab:stats_subjects}.
\begin{table}[htb]
	\centering
	
	\caption{Statistical data of the subjects in the SCA1 experiment. P is the number of participants. The $\mu()$ and the $\sigma()$ operators are the average and the standard deviation operators, respectively. Although being left-handed or right-handed does not make any difference due to the usage of an ambidextrous joystick, this information is presented for a general overview of the subjects.}
	\label{tab:stats_subjects}
	\begin{threeparttable}
		\centering
		\begin{tabular}{llllll}
			\toprule
			Scenario & P & F & $\mu(\text{Age})$ & $\sigma(\text{Age})$ & LH   \\ \hline
			\midrule
			S$_{\text{harsh}}$ & 15 & 1  &22.9  &3.6  &3      \\ \hline
			S$_{\text{mild}}$  & 3  & 0  &26.0    &5.2 & 0      \\ \hline
			\bottomrule
		\end{tabular}
		\begin{tablenotes}
			\small 
			\item F: female, LH: left-handed
		\end{tablenotes}
	\end{threeparttable}
\end{table}
\subsubsection{Pilot-model parameters}
The pilot-model in \eqref{e:perception_trigger}-\eqref{perception_variable2} includes statistical parameters $\mu_p$ and $\sigma_{p}$, the mean and the standard deviation of the time derivative of $c_i(t)$ (defined in \eqref{e:unscaled_rms_cfm}), respectively, and the parameters of the filter $G_1(s)$. The filter is chosen as $G_1(s)=\frac{2.25}{s^2+1.5s+2.25}$  so as to reflect the bandwidth of the pilot stick motion. To obtain the other statistical parameters, several flight simulations were run with the PD-control based autopilot in closed-loop and the resulting $c_i(t)$ values were calculated for 180s, both for the harsh and mild anomalies. The time-averaged statistics of the resulting profiles were used to calculate the statistical parameters as $\mu_{p}=0.028$, $\sigma_{p}=0.038$ for the harsh anomaly, and $\mu_{p}=0.091$, $\sigma_{p}=0.077$ for the mild anomaly.

The pilot model in \eqref{e:perception_trigger}-\eqref{perception_variable2} implies that the pilot signals the presence of the anomaly at the time instant when $K_t$ becomes unity, following which the control action switches from the autopilot to the human pilot. The action of the pilot based on this trigger is introduced in the experiment by choosing $t_s$, the instant of the sound signal, to coincide with the perception trigger. The corresponding $\Delta T$ is denoted as a "CfM-based" one. In order to benchmark this CfM-based switching action, two other switching mechanisms are introduced, one which we define to be "exact", where $t_s=t_a$, so that $\Delta T=0$, and another to be "late", where $\Delta T$ is chosen to be significantly larger than the CfM-based one. These choices are summarized in Table \ref{tab:timeline}.
\begin{table}[htb]
	\centering
	\caption{Timeline of Anomalies. Based on of the switching mechanism, the anomaly is reported to the subject with a sound signal at $t_{s}$. $\Delta T$, the alert time, is defined as the time elapsed between the sound signal and the occurrence of anomaly.} 
	\label{tab:timeline}
	\begin{threeparttable}
		\begin{tabular}{cc}
			\begin{minipage}{.5\linewidth}
				\centering
				\begin{tabular}{llll}
					\toprule
					Switch & $t_{a}[s]$  &$t_{s}[s]$  & $\Delta T [s]$  \\ \hline
					\midrule
					Late & 50     &55.5  & 5.5    \\ \hline
					Exact& 50     &50  & 0    \\ \hline
					CfM-based& 50  &51.1  & 1.1    \\ \hline
					\bottomrule
				\end{tabular}
				\begin{tablenotes}
					\small 
					\item $\enspace \enspace \enspace \enspace \enspace$ Scenario with the harsh anomaly, $S_{\text{harsh}}$
				\end{tablenotes}
			\end{minipage} &
			
			\begin{minipage}{.5\linewidth}
				\centering
				\begin{tabular}{llll}
					\toprule
					Switch  & $t_{a}[s]$ &$t_{s}[s]$  & $\Delta T [s]$   \\ \hline
					\midrule
					Late     &64  &74    &10    \\ \hline
					Exact    &64  &64    &0    \\ \hline
					CfM-based &64  &70.2  &6.2    \\ \hline
					\bottomrule
				\end{tabular}
				\begin{tablenotes}
					\small 
					\item $\enspace \enspace \enspace \enspace \enspace$ Scenario with the mild anomaly, $S_{\text{mild}}$
				\end{tablenotes}
			\end{minipage} 
		\end{tabular}
	\end{threeparttable}
\end{table}
\subsubsection{Results and Observations}
\subsubsection{Scenario 1: Harsh Anomaly}

We present the results related to the harsh anomaly defined in \eqref{e:harsh_transfer} for various alert times transmitted to the subjects. In order to compare the results obtained from the subjects, we also carried out numerical simulation results using the same alert times. The different cases are summarized in Table \ref{tab:explanations}.  The results obtained are summarized Table \ref{tab:general_harsh} using both the tracking error $e_{\text{rms}}$ and the corresponding CfM.
\begin{table}[htb]
	\centering
	\caption{Cases investigated with theoretical and experimental SCAs on the basis of different alert times. List of abbreviations used in the interpretation of the shared control architecture (SCA) human-in-the-loop simulation results.}
	\label{tab:explanations}
	\begin{tabular}{lll}
		\toprule
		Abbreviation   & Explanation    \\ \hline
		\midrule
		Auto & Autopilot simulation without control shift\\ \hline
		{Th$_{\text{late}}$} & SCA simulation with late alert time \\ \hline
		{Exp$_{\text{late}}$} & {SCA experiment with late alert time}\\ \hline
		{Th$_{\text{exact}}$} & {SCA simulation with exact alert time}\\ \hline
		{Exp$_{\text{exact}}$} &{SCA experiment with exact alert time}\\ \hline
		{Th$_{\text{CfM}}$} &{SCA simulation with CfM alert time}\\ \hline
		{Exp$_{\text{CfM}}$} &{SCA simulation with CfM alert time}\\ \hline
		\bottomrule			
	\end{tabular}
\end{table}	
All numbers reported in Table \ref{tab:general_harsh} are averaged over all 15 subjects. $e_{\text{rms}}$ was calculated using \eqref{rms-error} with $T_p=180$s, while CfM was calculated using \eqref{e:unscaled_cfm}- \eqref{e:cfm_desired} with $u_{\text{max}_{\text{el}}}=3$ and $\delta=0.25$. The statistical variations of both $e_{\text{rms}}$ and CfM over the 15 subjects are quantified for all three SCA experiments, for the late, exact, and CfM-based alert times are summarized in Table \ref{tab:stats_harsh}. We also calculate the average bumpless transfer metric $\rho$ and its standard error $\sigma_M(\rho)$ for these three cases in Table \ref{tab:ave_bumpless_harsh}. Table $\ref{tab:ave_bumpless_harsh}$ also provides the average reaction times $t_{\text{RT}}$ and average total reaction times $t_{\text{TRT}}$ of the subjects. 
\begin{table}[htb]
	\centering
	\caption{Averaged $e_{\text{rms}}$ and CfM values for S$_{\text{harsh}}$. The autopilot shows the worst tracking error performance among theoretical and experimental shared control architectures. The reason for a high CfM amount for the autopilot is the inability to effectively use the actuators to accommodate the anomaly.}
	\label{tab:general_harsh}
	\begin{tabular}{llllllll}
		\toprule
		S$_{\text{harsh}}$   &   Auto& ${\text{Th}}_{\text{late}}$ & $\text{Exp}_{\text{late}}$ & $\text{Th}_{\text{exact}}$ & $\text{Exp}_{\text{exact}}$ & $\text{Th}_{\text{CfM-based}}$ & $\text{Exp}_{\text{CfM-based}}$  \\ \hline
		\midrule
		$e_{\text{rms}}$  &   478   &  363   & \textcolor{black}{$383$}  & 319& \textcolor{black}{$354$} & 318  & \textcolor{black}{$348$} \\ \hline
		CfM  &   8.92  &  7.93  &  \textcolor{black}{$6.84$} & 7.83 &  \textcolor{black}{$6.80$} & 7.83 & \textcolor{black}{$7.06$}   \\ \hline
		\bottomrule			
	\end{tabular}
\end{table}	
\begin{table}[htb]
	\centering
	
	\caption{Mean, $\mu$ and standard error, $\sigma_M=\sigma / \sqrt{n}$, where $\sigma$ is standard deviation and $n$ is the subject size, of $e_{\text{rms}}$ (on the left) and CfM (on the right), for a harsh anomaly.}
	\label{tab:stats_harsh}
	\begin{threeparttable}
		\begin{tabular}{cc}
			\centering
			\begin{minipage}{.4\linewidth}
				\centering
				\begin{tabular}{lll}
					\toprule
					Experiment  & $\mu$  & $\sigma_{M}$   \\ \hline
					\midrule
					$\text{Exp}_{\text{late}}$     &383    & 16    \\ \hline
					$\text{Exp}_{\text{exact}}$    &354    & 15   \\ \hline
					$\text{Exp}_{\text{CfM-based}}$   &348     & 14   \\ \hline
					\bottomrule
				\end{tabular}
				\begin{tablenotes}
					\small 
					\item $\enspace \enspace \enspace$ $e_{\text{rms}}$
				\end{tablenotes}
				
			\end{minipage} &
			
			\begin{minipage}{.4\linewidth}
				\centering
				\begin{tabular}{lll}
					\toprule
					Experiment  & $\mu$  & $\sigma_{M}$   \\ \hline
					\midrule
					$\text{Exp}_{\text{late}}$     &6.84    &0.10    \\ \hline
					$\text{Exp}_{\text{exact}}$    &6.80    &0.12    \\ \hline
					$\text{Exp}_{\text{CfM-based}}$   &7.06   &0.11    \\ \hline
					\bottomrule
				\end{tabular}
				\begin{tablenotes}
					\small 
					\item $\enspace$ CfM
				\end{tablenotes}
			\end{minipage} 
		\end{tabular}
	\end{threeparttable}
\end{table}
\begin{table}[htb]
	\centering
	\caption{Averaged bumpless transfer metric, $\rho$, standard error, $\sigma_M(\rho)=\sigma(\rho) / \sqrt{n}$, where $\sigma$ is standard deviation of $\rho$ and $n$ is the subject size, and averaged reaction times $t_{\text{RT}}$ and total reaction times $t_{\text{TRT}}$ for S$_{\text{harsh}}$. The least amount of bumpless transfer of control happens to be in the case of CfM-based shared control architecture.}
	\label{tab:ave_bumpless_harsh}
	\begin{threeparttable}
		\begin{tabular}{lllll}
			\toprule
			Switch  & $\rho$ & $\sigma_M(\rho)$	& $t_{\text{RT}}$  [s] & $t_{\text{TRT}}$ [s] \\ \hline	
			\midrule
			\text{Late} &\textcolor{black}{216}	& 4.27 & 1.07    & 6.64     \\ \hline 
			\text{Exact}& \textcolor{black}{82}	& 1.37 & 0.98    & 0.98     \\ \hline 
			\text{CfM-based}& \textcolor{black}{26}	& 0.72 & 0.99  & 2.12     \\ \hline 	
			\bottomrule
		\end{tabular}
	\end{threeparttable}
\end{table}
\subsubsection{Observations}

\begin{figure}[htb]
	\centering
	\includegraphics[scale=0.55]{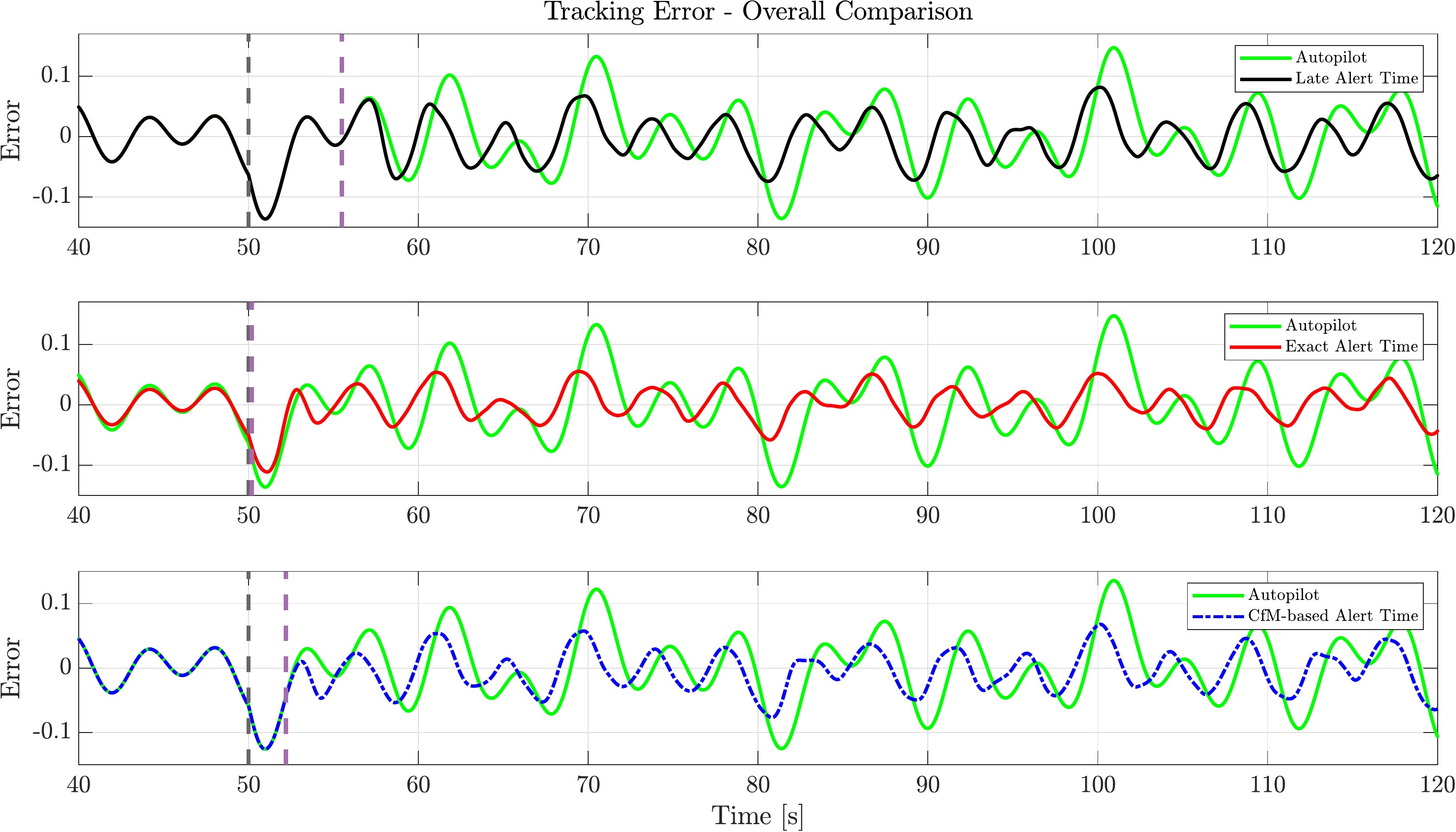}
	\caption{Comparison of tracking errors between the autopilot and all the alert timing mechanisms. The autopilot only case is the worst performer among all alert timing mechanisms which may be due to having predefined and fixed gains.}
	\label{fig:error_tracking}
\end{figure}

\begin{figure}[htb]
	\centering
	\includegraphics[scale=0.55]{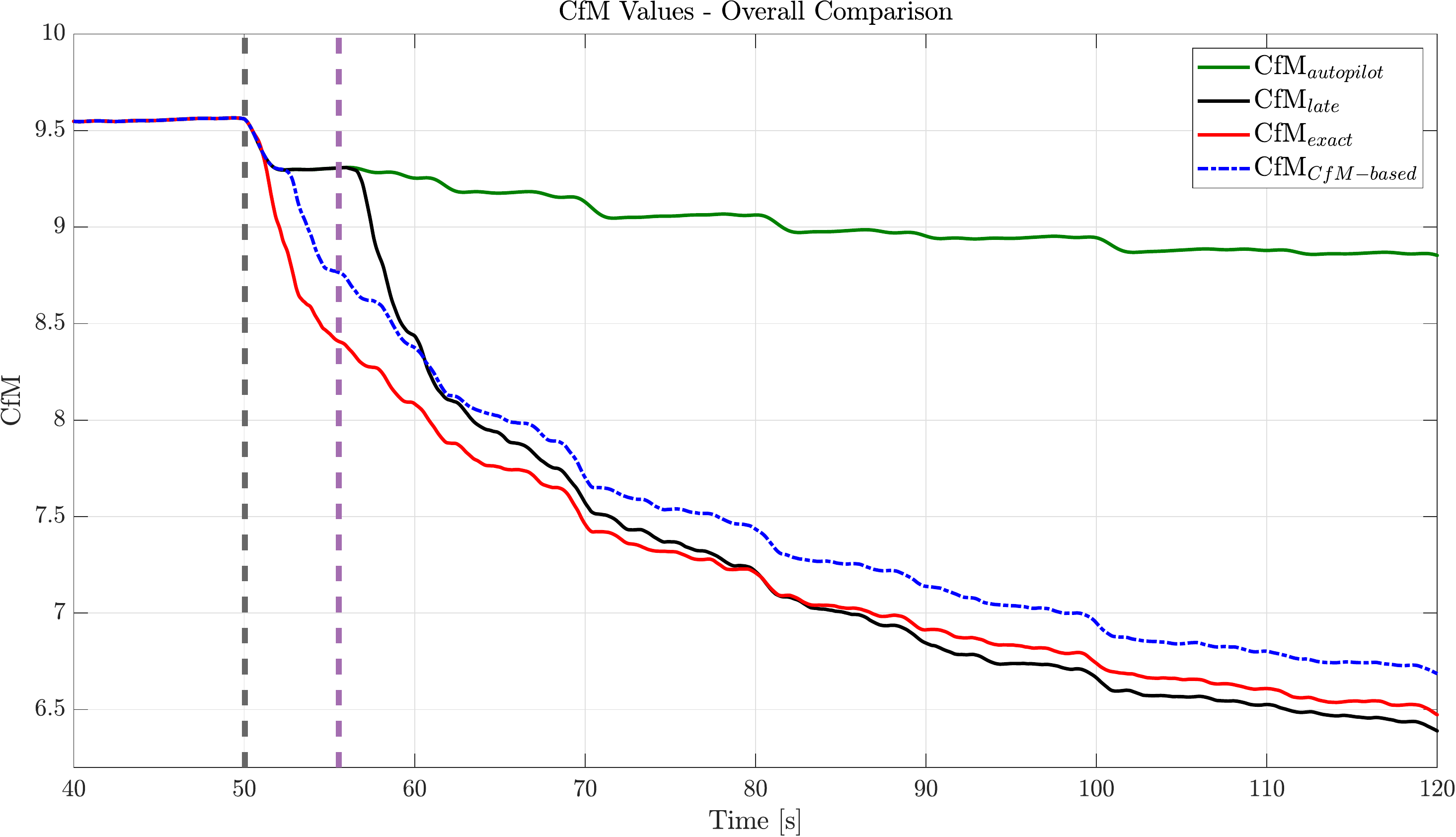}
	\caption{CfM variation for the autopilot and all the alert timing mechanisms. $\text{CfM}_{\text{late}}$ and $\text{CfM}_{\text{exact}}$ show changing trends during the simulation but $\text{CfM}_{\text{CfM-based}}$ prominently stays at the top among its counterparts especially after the transient effects of the anomaly ($t > t_{a}+10$).}
	\label{fig:CfM_comparison}
\end{figure}

The first observation from Table \ref{tab:general_harsh} is that the $e_\text{rms}$ for the CfM-based case is at least 25$\%$ smaller than the case with the autopilot alone. The second observation is that among the shared control architecture experiments, the one with the CfM-based alert time has the smallest tracking error while the one with the late alert time has the largest tracking error. However, it is noted that the difference between the exact and CfM-based cases is not significant. These two observations can also be confirmed from Figure $\ref{fig:error_tracking}$, where the tracking error is shown for different types of control structures including the autopilot, the late alert time SCA, exact alert time SCA and the CfM-based alert time SCA. The figure is obtained by averaging all of the experimental results. 
The third observation from Table \ref{tab:general_harsh} is that CfM based control switching from the autopilot to the pilot not only provides the smallest tracking error, but also the largest CfM value compared to other switching strategies. This can also be observed in Figure $\ref{fig:CfM_comparison}$, where CfM values, averaged over all subjects, are provided for different alert times. As noted earlier, the large CfM value of the autopilot results from inefficient use of the actuators which manifests itself with a large tracking error. The final observation, which comes from Table $\ref{tab:ave_bumpless_harsh}$, is that among different alert times, the one with the CfM-based alert time provides the smoothest transfer of control, with a bumpless transfer metric of $\rho=26$, while the late-alert provides $\rho=216$ and the exact alert gives $\rho=82$.

These observations imply the following: The pilot re-engagement after an anomaly should not be delayed until it becomes too late, which corresponds to a late-alert switching strategy. At the same time, immediate pilot action right after the anomaly detection might not be necessary either, which is indicated by the fact that the bumpless transfer metric and the CfM values for the CfM-based switching case is better than the exact-switching case. Instead, monitoring the CfM information carefully may be the appropriate trigger for the pilot to take over.
\subsubsection{Scenario 2: Mild Anomaly}
The averaged $e_{\text{rms}}$ (scaled by $10^4$) and CfM values for this case are given in Table $\ref{tab:general_mild}$. The averaged bumpless transfer metric $\rho$, its standard error $\sigma_M(\rho)$, the average reaction times $t_{\text{RT}}$ and the average total reaction times $t_{\text{TRT}}$ are presented in Table $\ref{tab:ave_bumpless_mild}$. The statistical variations of these metrics over the 3 subjects are shown in Table $\ref{tab:stats_mild}$.
\begin{table}[htb]
	\centering	
	\caption{Averaged $e_{\text{rms}}$ and CfM values for S$_{\text{mild}}$. The autopilot still shows the worst performance yet, this time, the error introduced is less than the one of the harsh anomaly.}
	\label{tab:general_mild}
	\begin{tabular}{llllllll}
		\toprule
		S$_{\text{mild}}$   &   Auto& ${\text{Th}}_{\text{late}}$ & $\text{Exp}_{\text{late}}$ & $\text{Th}_{\text{exact}}$ & $\text{Exp}_{\text{exact}}$ & $\text{Th}_{\text{CfM-based}}$ & $\text{Exp}_{\text{CfM-based}}$  \\ \hline
		\midrule
		$e_{\text{rms}}$   &   408     &  281   &\textcolor{black}{$259$}  & 297 & \textcolor{black}{$236$} & 243  & \textcolor{black}{$217$} \\ \hline
		\text{CfM}	&   8.78  &  7.90  &  \textcolor{black}{$7.75$} &   7.64 &  \textcolor{black}{$7.55$} & 8.13 & \textcolor{black}{$7.79$}   \\ \hline
		\bottomrule			
	\end{tabular}
\end{table}
\begin{table}[htb]
	\centering	
	\caption{Averaged bumpless transfer metric $\rho$; standard error  $\sigma_M(\rho)=\sigma(\rho) / \sqrt{n}$, where $\sigma$ is standard deviation of $\rho$, and $n$ is the subject size; averaged reaction time $t_{\text{RT}}$; and total reaction time $t_{\text{TRT}}$ for the experiment with a mild anomaly,  S$_{\text{mild}}$. The nature of the anomaly has a considerable effect on the bumpless transfer metric, that is, the difference between the exact and CfM-based alert timings is not readily noticeable as the one of the harsh anomaly.}
	\label{tab:ave_bumpless_mild}
	\begin{tabular}{lllll}
		\toprule
		Switch & $\rho$ & $\sigma_M(\rho)$	&$t_{\text{RT}}$ [s] & $t_{\text{TRT}}$ [s] \\ \hline
		\midrule	
		\text{Late} & 196 & 3.23	& 1.06    & 11.06     \\ \hline 
		\text{Exact}& \textcolor{black}{209}	& 8.88 & \textcolor{black}{1.02}    & \textcolor{black}{1.02}     \\ \hline 
		\text{CfM-based}& 203 & 2.87	& 0.95   & 7.20     \\ \hline 
		\bottomrule		
	\end{tabular}
\end{table}
\begin{table}[htb]   
	\centering
	\caption{Mean, $\mu$ and standard error, $\sigma_M=\sigma / \sqrt{n}$, where $\sigma$ is the standard deviation and $n$ is the subject size, of $e_{\text{rms}}$ (on the left) and CfM (on the left), for Scenario 2 with a mild anomaly.}
	\label{tab:stats_mild}
	\begin{threeparttable}
		\begin{tabular}{cc}
			\begin{minipage}{.4\linewidth}
				\centering
				\begin{tabular}{lll}
					\toprule
					Experiment  & $\mu$  & $\sigma_{M}$   \\ \hline
					\midrule
					$\text{Exp}_{\text{late}}$     &259       &19    \\ \hline
					$\text{Exp}_{\text{exact}}$    &236      &21    \\ \hline
					$\text{Exp}_{\text{CfM-based}}$   &218       &14    \\ \hline
					\bottomrule
				\end{tabular}
				\begin{tablenotes}
					\small
					\item $\enspace \enspace \enspace$ $e_{\text{rms}}$
				\end{tablenotes}
				
			\end{minipage} &
			
			\begin{minipage}{.4\linewidth}
				\centering
				\begin{tabular}{lll}
					\toprule
					Experiment  & $\mu$  & $\sigma_{M}$   \\ \hline
					\midrule
					$\text{Exp}_{\text{late}}$     &7.75    &0.15    \\ \hline
					$\text{Exp}_{\text{exact}}$    &7.55   &0.17    \\ \hline
					$\text{Exp}_{\text{CfM-based}}$   &7.61 &0.19    \\ \hline
					\bottomrule
				\end{tabular}
				\begin{tablenotes}
					\small
					\item $\enspace$ CfM
				\end{tablenotes} 
			\end{minipage} 
		\end{tabular}
	\end{threeparttable}
\end{table}
\subsubsection{Observations}
Similar to the harsh anomaly case, pure autopilot control results in the largest tracking error in the case of a mild anomaly. Also similar to the harsh anomaly case, pilot engagement based on CfM information produces the smallest tracking error, although the difference between the exact switching and CfM-based switching is not significant. However, in terms of preserving CfM, Table~\ref{tab:stats_mild} shows that no significant differences between different switching times can be detected, due to wide spread (high standard error) of the results. The same conclusion can be drawn for the bumpless transfer metric, although late switching has slightly smaller metric compared to the CfM-based switching. Again, the results are close to each other due to wide-spread. One reason for this can be low sample size, 3, in this experiment. One conclusion that can be drawn from these results is that although CfM based switching shows smaller tracking errors compared to the alternatives, the advantage of the proposed SCA in the presence of mild anomaly is not as prominent as in the case of harsh anomaly.   

%All of the observations in the harsh anomaly case are true here as well with the only exception that the $e_{rms}$ for the experimental case with CFM-based alert was smaller than its theoretical counterpart. This may be due to the fact that the subjects could mitigate the effects of the not so severe anomaly and outperform the theoretical case. It was also observed that compared to the harsh anomaly case, the corresponding CfM values were all uniformly higher. This is entirely reasonable as a harsh anomaly requires more capacity to be consumed in comparison to a mild one.

\subsection{Validation of Shared  Control  Architecture 2}

Unlike the SCA1, where the pilot took over control from the autopilot when an anomaly occurred, in SCA2, the pilot plays more of an advisory role, directing the autopilot that remains operational throughout. In particular, the pilot provides appropriate values $\mu$ and sometimes $\hat \Lambda_{f_p}$  as well. As in the previous section, we describe the validation using the same steps, starting from the experimental setup and ending with results and observations.

\subsubsection{Experimental Setup}

The same pilot joystick (Logitech  Extreme  3D  Pro  Joystick) used in validating SCA1 is employed for the SCA2 experiments. However, contrary to the SCA1 case, here the subjects uses the joystick only to enter the $\mu$ and $\hat \Lambda_{f_p}$ values using the joystick lever, normally used to simulate the throttle input. Joystick is not used for controlling the aircraft, since the autopilot is always in control. 

The aircraft model used in these human in the loop simulations is a nonlinear F-16 model, the details of which can be found at \cite{stevens2015aircraft} and \cite{nguyen1979simulator}.

\begin{figure}[htb]
	\centering
	\includegraphics[width=18cm]{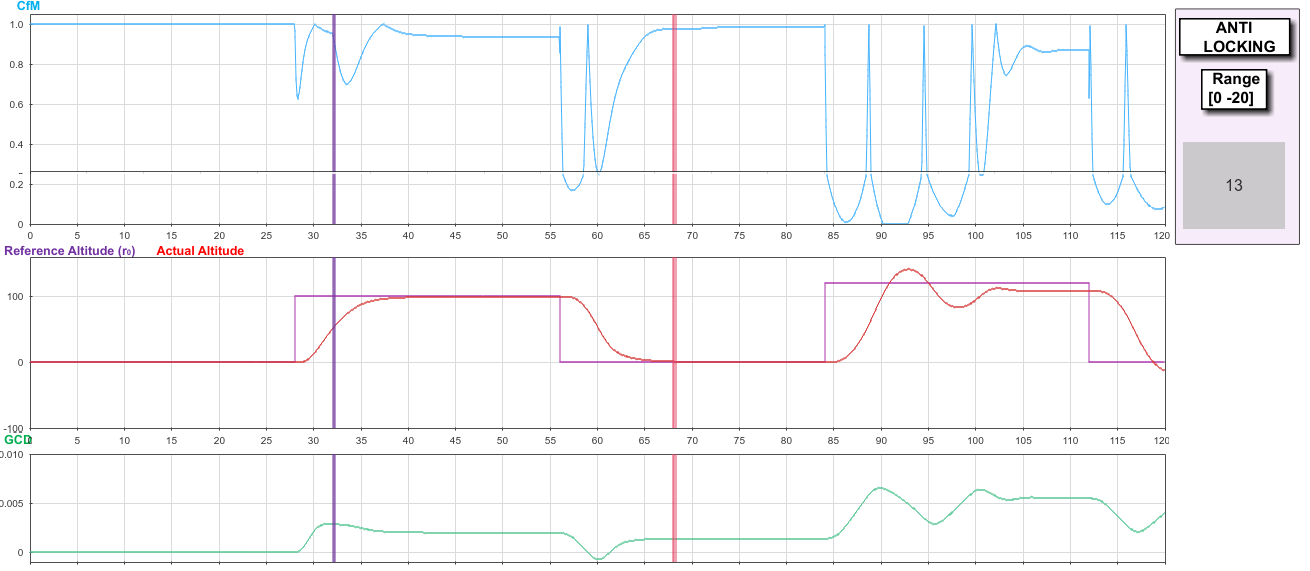}
	
	\caption{Flight screen interface for Shared Control Architecture 2. The subfigure at the top shows the variation of CfM during the flight. The vertical lines at $t_{a_1}=32$s and $t_{a_2}=68$s shows the instants of anomaly introduction.  
		The horizontal black line is the anti-locking border below which the $\mu$ input becomes effective. It is explained to the subjects, as well as demonstrated during training, that setting $\mu$ to high values where CfM variation is over this border has no influence on CfM, which is apparent from (\ref{e:u_i_2}). The subfigure in the middle shows reference tracking and the one in the bottom is the time variation of the graceful command degradation during flight.}
	\label{fig:simscreen}
\end{figure}

The flight screen that the subjects see is shown in Figure~\ref{fig:simscreen}. There are three subplots, which are CfM variation (top), %($c_{\text{el}}(t)$) in (\ref{e:unscaled_rms_cfm}), ($b$) 
reference command ($r_0$) tracking (middle) 
%\& altitude change ($h$) and ($c$) 
and the evolution of graceful command degradation (GCD) (bottom). The horizontal black line in the CfM variation subplot corresponds to the upper bound of the virtual buffer, $[0, \enspace \delta]=[0, \enspace 0.25]$. The small rectangle at the upper right serves the purpose of showing the amount of the $\mu$ input, entered via the joystick lever. In this rectangle, the title "Anti Locking" is used to emphasize the purpose of the $\mu$ input, which is preventing the saturation/locking of the actuators. There is also another region in this rectangle called the ``Range", which shows the limits of this input. 
In the snapshot of the pilot screen shown in Figure~\ref{fig:simscreen},
a scenario with two anomalies introduced at $t_{a_{1}}=32$s and  $t_{a_{2}}=68$s is shown. The instant of anomaly occurrences are marked with vertical lines, the colors of which indicate the severity of the anomaly. The subjects are trained to understand and respond to the severity and the effect of the anomalies by monitoring the colors, CfM information and the tracking performance. The details of subject training are provided in the ``Experimental Procedure" section.

\subsubsection{Anomaly}

The anomaly considered in this experiment is loss of actuator effectiveness indicated by $\Lambda_{f}$ in (\ref{e:plant}). $\Lambda_{f}$ is a $(2 \times 2)$ diagonal matrix with equal entries that are between 0 and 1, where 0 corresponds to complete actuator failure and 1 corresponds to no failure. In the experiments, two anomalies are introduced at times $t=t_{a_1}$ and $t=t_{a_2}$. Consequently, the diagonal entries of $\Lambda_{f}$ vary as 
\begin{equation}\label{e:anomalies_sca2}
\lambda_{f_{i}}=
\left\{\begin{array}{ll}
1, & t< t_{a_{1}}\\
0<\lambda_{f_{1}}<1, & t_{a_{1}} \leq t < t_{a_{2}}\\
0<\lambda_{f_{2}}<1, & t_{a_{2}} \leq t
\end{array}\right.
\end{equation}
where $\lambda_{f_{i}}$ refers to the diagonal entries for the $i^{th}$ anomaly introduction. The anomaly injections are communicated to the subjects with colored vertical lines appearing on the pilot screen, as shown in Figure~\ref{fig:simscreen}, together with sound alerts. Three different anomalies with different severities are used during the experiments. The anomaly severity quantities, corresponding severity labels and color codes are provided in Table~\ref{tab:colorcodes}.

\begin{table}[htb]
	\centering
	
	\caption{Anomaly severities. 
		%The subjects are given a set of different scenarios in which these severities are either separately applied or associated. 
		The anomalies are introduced in an audiovisual fashion, that is, the subjects perceive the anomaly by a specific sound and also have the chance to recognize it by its corresponding color. 
		%They are exposed to each type of anomaly and their combined forms to decide on a plausible $\mu$ input. 
		Times of anomaly injections are marked as a vertical line which appears at the instant of the anomaly, after which the subject responds by providing an appropriate $\mu$ value.}
	\label{tab:colorcodes}
	\begin{tabular}{lll}
		\toprule
		$\Lambda_{f}$	& Severity  & Color Code   \\ \hline
		\midrule
		$0.30$	& Low    &  \enspace \enspace \colorbox{green}{\textcolor{green}{OOO}}       \\ \hline
		$0.20$	& Middle &  \enspace \enspace \colorbox{violet}{\textcolor{violet}{OOO}}       \\ \hline
		$0.15$  & High   &  \enspace \enspace \colorbox{purple}{\textcolor{purple}{OOO}}          \\ \hline
		
		\bottomrule
	\end{tabular}
\end{table}

\subsubsection{Experimental Procedure}

\begin{figure}[htb]
	\centering
	\includegraphics[scale=1]{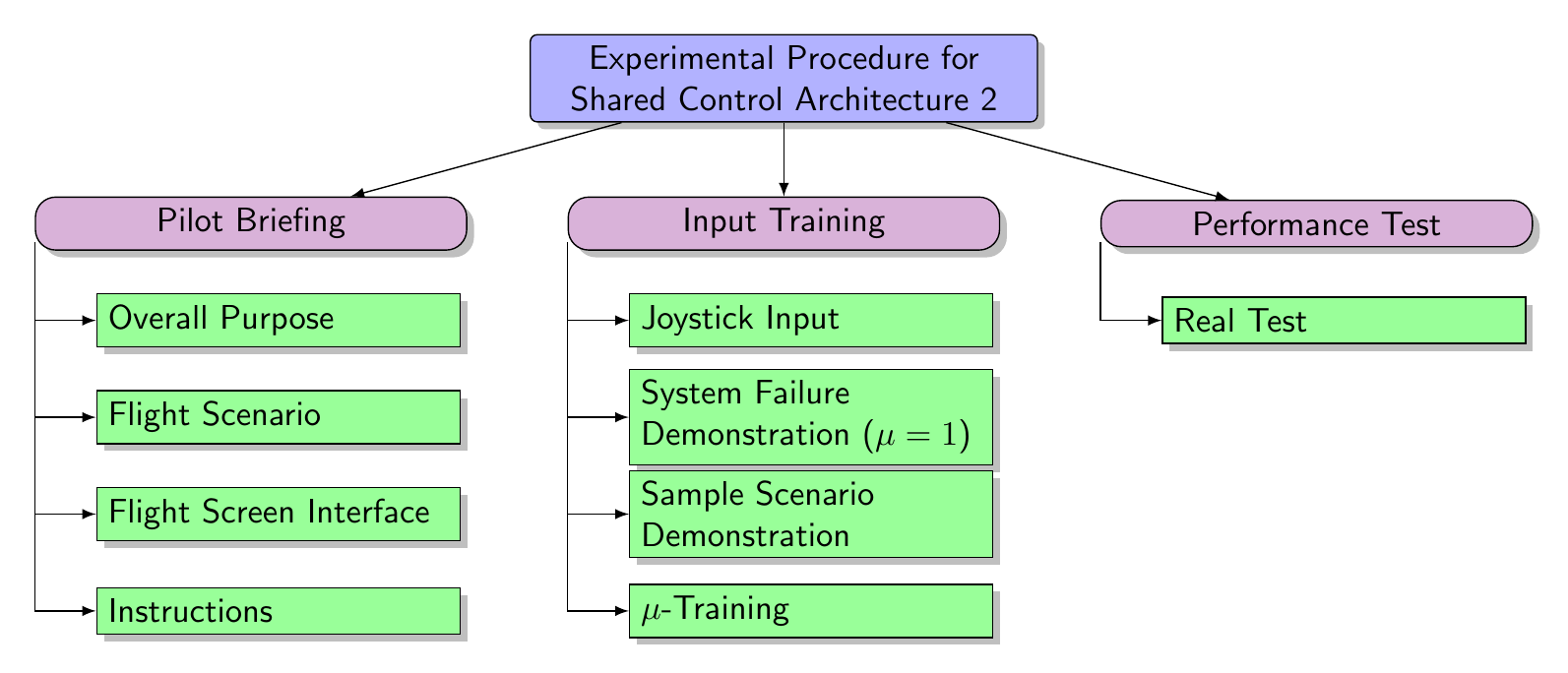}
	\caption{Experimental procedure breakdown for the Shared Control Architecture 2. Three main tasks constituting the procedure are observed. In \textit{Pilot Briefing}, the subjects read the pilot briefing, review it with the experiment designer and have a question and answer session. In \textit{Input Training}, the subjects learn how to provide the auxiliary inputs to the autopilot using the joystick. They also  undergo a $\mu$-input training, the details of which are covered in Section ``Experimental Procedure''. Finally, in \textit{Performance Test}, a real test with two successive anomalies $\Lambda_{f_i}$ is conducted.}
	\label{fig:Breakdown_Structure_SCA2}
\end{figure}

As in SCA1, the experimental procedure consists of three parts, \textit{Pilot Briefing}, \textit{Input Training} and \textit{Performance Test}. (See Figure \ref{fig:Breakdown_Structure_SCA2} for a schematic).

The first part of the procedure is the \textit{Pilot Briefing}, in which the subjects are demanded to read a pilot briefing to acquire some prior knowledge about the experiment. The briefing consists of four main sections, namely, \textit{Overall Purpose}, \textit{Flight Scenario}, \textit{Flight Screen Interface} and \textit{Instructions}. In these sections, the main concepts and the experimental setup are covered by the experiment designer. The scaled toy aircraft prototype described in the previous experiment for SCA1 is utilized in this experiment as well, especially in conjunction with the description of control surfaces in the aircraft, since the anomalies are related to actuator failures.
%
%
% Sidebar Comment:
% The scaled toy aircraft prototype described in the previous experiment for SCA1 is utilized in this experiment as well, especially in conjunction with the description of control surfaces in the aircraft, since the anomalies are related to actuator failures. The details about this pilot briefing can be found in Sidebar "\nameref{sidebar-HITLS2}".
% 
%
%
% Sidebar Comment
% \caption{Experimental procedure breakdown for the Shared Control Architecture 2. Three main tasks constituting the procedure are observed. In \textit{Pilot Briefing}, the subjects read the pilot briefing, the details of which are given in Sidebar ``\nameref{sidebar-HITLS2}", review it with the experiment designer and have a question and answer session. In \textit{Input Training}, the subjects learn how to provide the auxiliary inputs to the autopilot using the joystick. They also  undergo a $\mu$-input training, the details of which are covered in Section ``Experimental Procedure''. Finally, in \textit{Performance Test}, a real test with two successive anomalies $\Lambda_{f_i}$ is conducted.}
%
The second part is the \textit{Input Training}, in which the subjects are first introduced to the joystick lever, which they are required to use to enter the $\mu$ input (See Figure \ref{fig:collage_experiments}). By properly moving the lever, an integer value of $\mu$, ranging from $1$ to $20$, can be given to the controller. Following this, a demonstration test is conducted by the experiment designer. In this test, a sample scenario with two anomalies is run, where the input $\mu$ is fixed to its nominal value of $\mu=1$, throughout the flight. It is explicitly shown that 1) the actuators reaches their saturation limit, and thus the CfM becomes zero, many times during the flight, which jeopardizes the aircraft stability, 2) it takes for the altitude, $h$, a long time to recover to follow the reference command, and 3) a certain graceful command degradation occurs to relax the performance goals. While running the simulation, the elements in the flight screen interface are also explained. 

Following the demonstration test, another sample scenario is run by the designer, in which suitable $\mu$ values are provided upon occurrence of the anomalies. Different from the previous demonstration, it is pointed out that 1) with a proper $\mu$, CfM can be kept away from zero and 2) GCD is kept minimal. By this demonstration, the subjects are expected to appreciate that suitable $\mu$ inputs help the autopilot recover from severe anomalies in an efficient and swift manner.

Finally, a $\mu$-input-training is performed on the subjects. Six scenarios are introduced to the participants in an interactive manner, by which they learn how to be involved in the overall control architecture. 
%These scenarios encompass three single (a different $\Lambda_{f_{i}}$ in each) and three double (combination of two different $\Lambda_{f_{i}}$ in each) anomalies in the simulation. During the experiments, three anomalies of different severities, $\Lambda_{f_{i}}$, are considered and color-coded as presented in Table $\ref{tab:colorcodes}$. 
%
%In the scope of coming up with an optimal input, they are trained to give a suitable $\mu$ value upon occurrence of an anomaly. 
As a first step, three scenarios with single anomalies are considered. In each severity of the anomaly, an optimal $\mu$, which trades off CfM with GCD, is conveyed to the subjects. Upon doing this, attention is drawn to the fact that each anomaly causes a certain sharp drop in the CfM variation. In other words, more drastic drops occur in CfM with the increase in the severity of anomaly. Following this, three other scenarios with two successive anomalies are studied. It is noted by the combination of different anomalies that nonlinear effects are present in the flight simulation; that is, the $\mu$ values corresponding to single anomalies are no longer effective when the anomalies are combined. It is noted that in both the training and the performance tests the anomaly severity estimation $\hat \Lambda_{f_p}$ input is automatically fed to the controller with an error of 0.2 in absolute value. Then, to compare the effect of this estimation input, we compare two pilot types where the estimation is not provided in one case and provided in the other. 

The third part is the \textit{Performance Test}, in which the subjects are expected to handle a combination of a highly severe anomalies ($\Lambda_{f_{1}}=0.20$ and $\Lambda_{f_{2}}=0.15$) in a flight simulation. They are expected to give suitable $\mu$ values on the basis of the training they obtain. Also, in both the training and the performance tests, the introduction of anomaly times are chosen to be completely random to prevent the subjects from making a guess whether or not the anomaly is about to happen.

\begin{figure}[htb]
	\centering
	\includegraphics[scale=0.70]{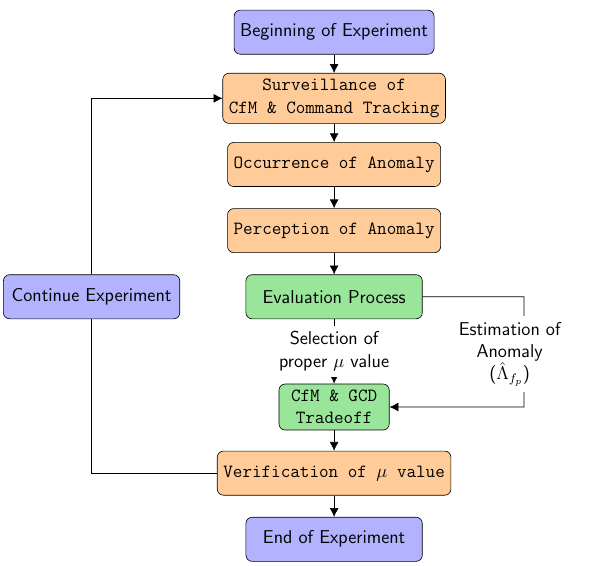}	
	\caption{Algorithm of pilot tasks. This presents the step-by-step procedure which should be followed by the subject.}	
	\label{fig:flowchart}
\end{figure}

The set of tasks to be tackled by the subjects is summarized as a flowchart in Figure $\ref{fig:flowchart}$. These tasks are explained interactively during the sample scenario demonstration (see Figure \ref{fig:Breakdown_Structure_SCA2}), where the subjects have the opportunity to practise the steps in controlling the flight simulation. 
\subsubsection{Details of the Human Subjects}

The experiment was performed by 10 subjects all of whom attended the previous experiment. This choice was made deliberately since the subjects of the previous experiment had an acquaintance with the autopilot and shared control concepts. For this reason, the pilot briefing regarding this experiment was written in a tone that the subjects were already familiar with these basic notions.  The experiment was approved by the Bilkent University Ethics Committee and an informed consent was taken from each subject. Some statistical data pertaining to the subjects are given in Table \ref{tab:stats_subjects_mu_exp}.
%
% Sidebar Comment:
% The experiment was performed by 10 subjects all of whom attended the previous experiment. This choice was made deliberately since the subjects of the previous experiment had an acquaintance with the autopilot and shared control concepts. For this reason, the pilot briefing (see Sidebar ``\nameref{sidebar-HITLS2}'') regarding this experiment was written in a tone that the subjects were already familiar with these basic notions.  The experiment was approved by the Bilkent University Ethics Committee and an informed consent was taken from each subject. Some statistical data pertaining to the subjects are given in Table \ref{tab:stats_subjects_mu_exp}.
%
%
\begin{table}[htb]
	\centering
	
	\caption{Statistical data of the subjects in the SCA2 experiment. P is the number of participants. The $\mu()$ and the $\sigma()$ operators are the average and the standard deviation operators, respectively.}
	\label{tab:stats_subjects_mu_exp}
	\begin{threeparttable}
		\centering
		\begin{tabular}{lllll}
			\toprule
			Scenario &P &F & $\mu(\text{Age})$ & $\sigma(\text{Age})$    \\ \hline
			\midrule
			Real Test &10 &0 &24.2   &1.8      \\ \hline
			\bottomrule
		\end{tabular}
		\begin{tablenotes}
			\small 
			\item F: female
		\end{tablenotes}
	\end{threeparttable}
\end{table}

\subsubsection{Results and Observations}

\begin{figure}[htb]
	\centering
	\includegraphics[width=16.5cm]{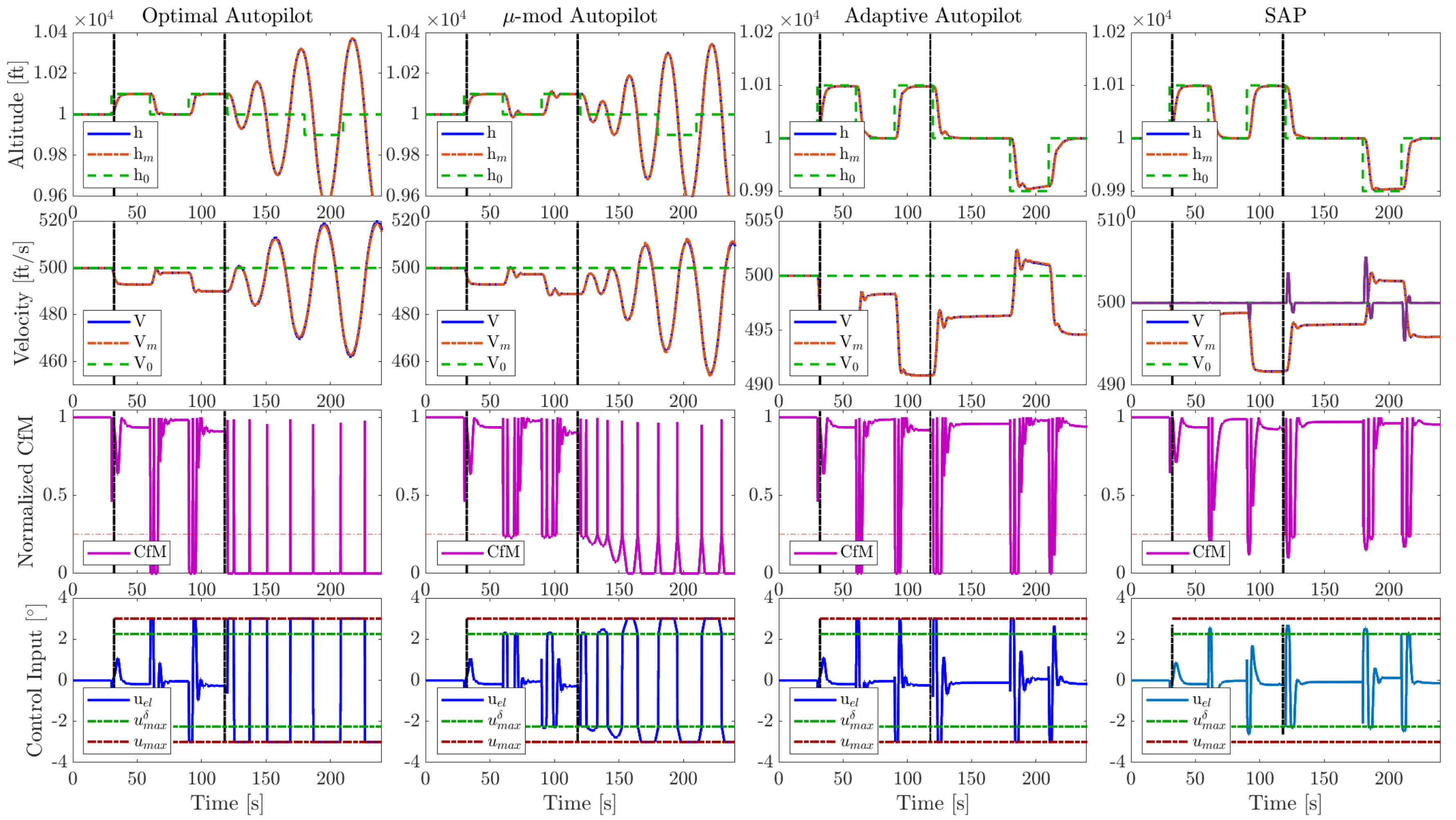}
	\caption{SCA2 vs autopilot-only cases. Each column presents the results of a specific controller whereas each row shows the comparison of these controllers based on altitude tracking $h$, velocity tracking $V$, CfM and elevator control input $\delta_{\text{el}}$, respectively. The horizontal dashed red line in the $3^{\text{rd}}$ row shows the buffer limit ($\delta=0.25$). The horizontal dashed red and green lines in the $4^{\text{th}}$ row show the limitations posed by $u_{\text{max}_{\text{el}}}=3$ and $u_{\text{max}_{\text{el}}}^{\delta}=2.25$.}
	\label{fig:subplots16}
\end{figure}

The results with 10 subjects inputting $\mu$ into the adaptive autopilot, are shown below. The autopilots consist of an adaptive controller as in \cite{farjadian2018resilient}, which has no apparent interface with the autopilot, a $\mu$-mod adaptive controller with a fixed value of $\mu=50$, and an optimal controller. The numerical results averaged over 10 subjects are presented in Table $\ref{tab:scaversus}$. The indices in curly brackets $\{h,v\}$ denote the altitude and velocity, respectively. SAP and SUP refer to the ``Situation Aware Pilot'' and the ``Situation Unaware Pilot'', respectively. SUP provides only a $\mu$ input and SAP provides both $\mu$ and $\hat \Lambda_{f_{p}}$ to the autopilot. The results for the SUP are obtained by simulating the performance tests by using only the $\mu$ inputs provided by the subjects, without their severity estimation inputs $\hat \Lambda_{f_p}$.  Therefore, to obtain the SUP results, the equation set \eqref{e:matching} is used, instead of \eqref{e:matching2}, while still incorporating the same mu values that were entered by the participants. Furthermore, the tracking performance, $\gamma_i$, is calculated as 
\begin{equation}\label{eqn_rho_rmse}
\gamma_{i} = \text{RMSE}_{i}^{+}-\text{RMSE}_{i}^{-},
\end{equation}
where
\begin{equation}\label{eqn_rmse}
\text{RMSE}_{i}^{-}=\text{rms}(e_{i})|_{0}^{t_{a_{1}}}, \quad \text{RMSE}_{i}^{+}=\text{rms}(e_{i})|_{t_{a_{1}}}^{T}.
\end{equation}
\begin{table}[tb]
	\centering
	\begin{threeparttable}
		
		\caption{Shared Control Architecture (SCA) 2 versus autopilot-only cases. SCA, both with a ``Situation Aware Pilot'' (SAP) and a ``Situation Unaware Pilot'' (SUP), results in higher tracking performance $\gamma$, higher CfM values and smaller GCD. On the other hand, SAP performs better than SUP in all these metrics.}
		\label{tab:scaversus}
		\begin{tabular}{lllll}
			\toprule
			Method & RMSE$^{-}_{\{h,v\}}$	& $\gamma_{\{h,v\}}$  & CfM  & GCD    \\ \hline
			\midrule
			SAP & $\{60,23\}\times10^{-4}$	    & $\{36,135\}\times10^{-4}$  & 1.21  & 24.8$\times 10^{-4}$    \\ \hline
			SUP & $\{60,23\}\times10^{-4}$	    & $\{51,152\}\times10^{-4}$  & 1.16  & 24.8$\times 10^{-4}$    \\ \hline
			Adaptive&$\{24.1,0.4\}$   &$\{0.51,1.06\}$  & 0.92  & NA    \\ \hline
			$\mu$-mod&$\{60,23\}\times10^{-4}$  &$\{0.10, 0.27\}$  & 0.81 &  342$\times 10^{-4}$    \\ \hline
			Optimal& $\{24.1,0.4\}$   &$\{160.8, 13.8\}$  & 0.84 &  NA \\ \hline
			\bottomrule
		\end{tabular}
		\begin{tablenotes}
			\small 
			\item CfM: Capacity for Maneuver
			\item GCD: Graceful Command Degradation
			\item NA: Not Applicable
		\end{tablenotes}
	\end{threeparttable}
\end{table}
In Figure $\ref{fig:subplots16}$, a comprehensive comparison of the SCA2 with autopilot-only cases is given as a matrix of $4\times4$ plots. Each column presents the results of a specific controller whereas each row shows the comparison of these controllers based on altitude tracking $h$, velocity tracking $V$, CfM and elevator control input $\delta_{\text{el}}$, respectively. The horizontal dashed red line in the $3^{\text{rd}}$ row shows the buffer limit. The horizontal dashed red and green lines in the $4^{\text{th}}$ row show the limitations posed by $u_{\text{max}}$ and $u_{\text{max}}^{\delta}$ introduced in ($\ref{e:u_max}$).
The results given in Table~\ref{tab:scaversus} and Figure~\ref{fig:subplots16} can be summarized as follows: First, it is observed that SCA2, whether with SAP and SUP, outperforms all the other autopilot-only cases. SCA2 provides a higher tracking performance $\gamma$, a higher CfM and a lower GCD. Second, SAP shows a better performance  than other controllers, including SUP, throughout the simulation run by not only showing a higher tracking performance, but also preventing CfM from reaching zero (saturation point). Third, a quick inspection of the first row of Figure~\ref{fig:subplots16} shows that both optimal and $\mu-$mod autopilots fail to respond to the second anomaly. This can be explained by the fact that CfM reaches to zero (saturation point) many times, especially in the case of optimal controller. Fourth, the adaptive autopilot shows an acceptable performance up to the second anomaly, but demonstrates degraded behavior with repeated saturation. It is noted that the elevator in the case of adaptive autopilot also hits the saturation point many times and shows a more oscillatory response compared to SAP.

\begin{figure}[htb]
	\centering
	\includegraphics[width=16.5cm]{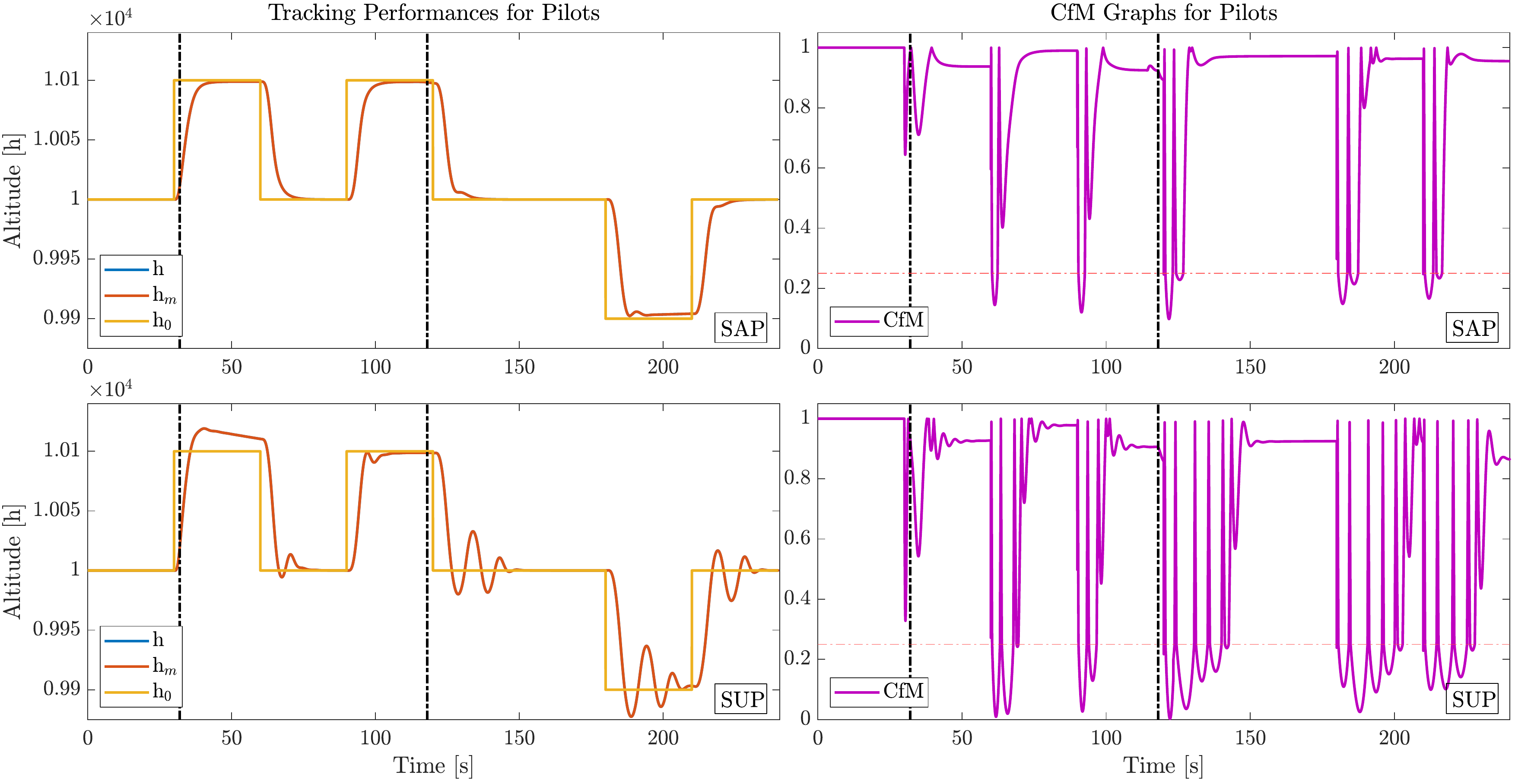}
	
	\caption{Comparison of the SAP and the SUP for the same scenario. The pure difference between these two pilots is based on the additional anomaly estimation inputn$\hat \Lambda_{f_{p}}$. It greatly attenuates the oscillatory tracking behavior around the reference model and contributes to a fast recovery from the buffer region.}
	\label{fig:supsap}
\end{figure}

To emphasize the effect of anomaly estimation input $\hat \Lambda_{f_{p}}$, the performances of SCA2 with SAP and SUP are shown separately in Figure~$\ref{fig:supsap}$. It is seen that SAP performs better than the SUP in terms of both tracking and CfM metrics. The selection of a suitable $\mu$ and the introduction of anomaly estimation, even with an error of 0.2, contribute drmatically to resilient performance, which can also be numerically verified by performance metrics given in Table $\ref{tab:scaversus}$.
\section{Summary and Future Work}

Although increased automation has made it easier to control aircraft, ensuring a safe interaction between the pilots and the autopilots is still a challenging problem, especially in the presence of severe anomalies. The domain of decision making is common to both human experts and feedback control systems. However, the process of detecting the anomaly, mitigating the anomaly, speed of response, intrinsic latencies, and the overall decision making procedure may vary between the pilot and the autopilot. Current approach consists of autopilot solutions that disengage themselves in the face of an anomaly and the pilot takes over. This may cause reengagement of the pilot at the worst possible time, which can result in undesired consequences, and a bumpy transfer. Our thesis in this paper is that a shared control architecture where the decision making of the human pilot is judiciously coordinated with that of an advanced autopilot is highly attractive in flight control problems where severe anomalies are present. We present two distinct architectures through which such coordination can take place, and validate these architectures through human-in-the-loop simulations. Though the type of coordination varies between these two architectures, both employ a common principle from cognitive engineering, namely CfM. While these architectures were proposed in recent publications \cite{farjadian2016resilient, thomsen2019shared, farjadian2018resilient}, a common framework within which to view them as well as their validation using human subject data are the main contributions of this paper.

In its most basic definition, CfM refers to the remaining capacity of the actuators that can be used for bringing the aircraft to safety. Based on this definition, CfM becomes small as the actuators get close to their saturation limits. The two architectures presented are denoted as SCA1 and SCA2. In SCA1, the pilot takes over the control from  the autopilot using the monitored CfM information. In SCA2, the pilot takes on the role of a  supervisor helping the autopilot whenever a need arises, based on the CfM information. Whenever the aircraft experiences a severe anomaly, the pilot assesses the situation based on their CfM monitoring and intervenes by providing two control system parameter estimates to the autopilot. This  increases  the  effectiveness  of  the  autopilot.  Using  human-in-the-loop  simulations,  it is shown that CfM based interactions, for both SCA1 and SCA2, provides smaller tracking errors and larger overall CfM.

In the experiments conducted, the subjects used included a commercial airline pilot and several university students, that are trained using a systematic procedure. As an experimental setup, a laptop screen is used as the pilot screen and a commercially available joystick is employed as an interface between the subjects and control system. The results reported here are based on this simple setup and can be considered as a first step towards understanding the effect of different interaction mechanisms  between pilots and autopilots. A large-scale study, with more subjects, with a range of piloting expertise, and with high-fidelity flight simulators are all the next set of topics that remain to be pursued to further validate the shared control concept.

As automation increases in engineered systems, creation of new cyber physical \& human systems is inevitable. There will be a variety of scenarios where humans and machines will need to interact and engage in combined decision making in order to lead to resilient autonomous behavior. These interactions will be complex, distinct, and will require new tools and methodologies.  Deeper engagement with the social science community so as to get better insight into human decision making and advanced modeling approaches is necessary. The results reported here should be viewed as a first of several steps in this research direction.

\section{Acknowledgment}
The first two authors would like to express their gratitude towards the support of The Scientific and Technological Research Council of Turkey under the grant number 118E202. The third author would like to gratefully acknowledge the support of Boeing through the Boeing Strategic University Initiative.

%\subsubsection{SAP versus SUP}

%A comparison between the SAP and the SUP, which is averaged over 10 subjects is given in Figure $\ref{fig:supsap}$. It is seen that SAP performs better than the SUP in terms of both tracking and CfM maximization. This conspicuously exhibits the beneficial effect of the anomaly  estimation input in response to the anomaly. The selection of suitable $\mu$ and even slightly faulty anomaly estimations ($|| \Delta \Lambda_{f_{p}}||=0.2$) contribute tremendously to resilient performance, which might also be numerically verified by the tracking performance metric, $\rho$, in Table $\ref{tab:scaversus}$.     

% ***** Experimental Results ******

% ***** References ******
\bibliographystyle{IEEEtran}
\bibliography{IEEEabrv,references}
% ***** References ******

\end{document}